\pdfoutput=1
\newcommand*{\ATLASLATEXPATH}{}
\documentclass[PAPER, cernpreprint, texlive=2016, UKenglish, LANGEDIT=true]{\ATLASLATEXPATH atlasdoc}
\LEcontact{Pat Ward cpw1@hep.phy.cam.ac.uk}
 
\usepackage{\ATLASLATEXPATH atlaspackage}
\usepackage{\ATLASLATEXPATH atlasbiblatex}
 
\usepackage{\ATLASLATEXPATH atlasphysics}
 
\graphicspath{{logos/}{figures/}}
 
\usepackage{ANA-HIGG-2017-14-PAPER-defs}

\AtlasTitle{Measurement of the production cross section for a Higgs boson in association with a vector boson in the $H \rightarrow WW^{\ast} \to \ell\nu\ell\nu$ channel in $pp$ collisions at $\sqrt{s}$ = 13~TeV with the ATLAS detector}

\AtlasAbstract{
A measurement of the Higgs boson production cross sections  via associated
$WH$ and $ZH$ production using $H \rightarrow WW^{\ast} \rightarrow
\ell\nu\ell\nu$ decays,  where $\ell$ stands for either an electron or a muon,  is presented.
Results for combined $WH$ and $ZH$ production are also presented.
The analysis uses events produced in proton--proton
collisions collected with the ATLAS detector at the Large Hadron Collider in 2015 and 2016.
The data correspond to an integrated luminosity of $36.1\,\mathrm{fb}^{-1}$ recorded at a centre-of-mass energy of 13~TeV.
The products of the $H \rightarrow WW^{\ast}$ branching fraction times the $WH$ and $ZH$ cross sections are measured to be
$0.67^{+0.31}_{-0.27}$(stat.)$^{+0.18}_{-0.14}$(syst.)~pb and $0.54^{+0.31}_{-0.24}$(stat.)$^{+0.15}_{-0.07}$(syst.)~pb respectively,
in agreement with the Standard Model predictions.
}

\AtlasRefCode{HIGG-2017-14}
 
\PreprintIdNumber{CERN-EP-2019-038}

\AtlasJournal{Phys.\ Lett.\ B.}
\AtlasJournalRef{\PLB 798 (2019) 134949}
\AtlasDOI{10.1016/j.physletb.2019.134949}

\AtlasCoverSupportingNote{Supporting Note}{https://cds.cern.ch/record/2295764}
 
\AtlasCoverCommentsDeadline{???}
 
\AtlasCoverAnalysisTeam{Antonio Baroncelli, Kathrin Becker, Claudia
Bertella, Michela Biglietti, Carsten Burgard, Dominik Duda, Ada
Farilla, Frank Fiedler, Frank Filthaut, Ralf Gugel, Yesenia Hernandez
Jimenez, Sana Ketabchi, Karsten Koeneke, Peter Krieger, Aaron Liblong,
Bruce Mellado, Sophio Pataraia, Christian Schmitt, Marco Sessa, Jonas
Strandberg, Marcel Weirich, Natalie Wieseotte}
 
\AtlasCoverEdBoardMember{Zhiqing~Zhang~(chair), Aaron Armbruster,
Andrea Bocci}

\AtlasCoverEgroupEditors{atlas-HIGG-2017-14-editors@cern.ch}
 
\AtlasCoverEgroupEdBoard{atlas-HIGG-2017-14-editorial-board@cern.ch}
 
\hypersetup{pdftitle={ATLAS document},pdfauthor={The ATLAS Collaboration}}
 
\DeclareUnicodeCharacter{2009}{FIXME}
 
\begin{document}
 
\maketitle

\section{Introduction}
 
Higgs boson production in association with a $W$ or $Z$  boson,
which is respectively denoted by \WH and \ZH, and collectively
referred to as \VH associated production in the following, provides
direct access to the Higgs boson couplings to weak bosons. In
particular, in the \WH mode with subsequent \HWW  decay,
the Higgs boson couples only to W bosons, at both the production and
decay vertices.
 
This paper presents a measurement of the corresponding production
cross sections through the decay \HWWlnln, using proton--proton collisions
at a centre-of-mass energy of $\sqrt{s} =$ 13~\TeV.
The data correspond to an integrated luminosity of $36.1$~fb$^{-1}$ and were
recorded by the ATLAS detector at the Large Hadron Collider (LHC).
Previous measurements at $\sqrt{s}=8$~\TeV\ were
performed by the ATLAS~\cite{HIGG-2013-15} and
CMS~\cite{CMS-HIG-13-023} Collaborations and recently at
$\sqrt{s}=13$~\TeV\ with $35.9\,\mathrm{fb}^{-1}$ of data by the CMS
Collaboration~\cite{CMS-HIG-16-042}. Recent results at  $\sqrt{s}=13$~\TeV\ on
\VH production in other decay modes 
can be found in Refs.~\cite{HIGG-2016-21, HIGG-2016-22, HIGG-2018-04,HIGG-2017-01,CMS-HIG-18-016,CMS-HIG-18-007}.
 
\begin{figure}[htb]
\centering
\subfloat[] {
\includegraphics[width=0.3\linewidth]{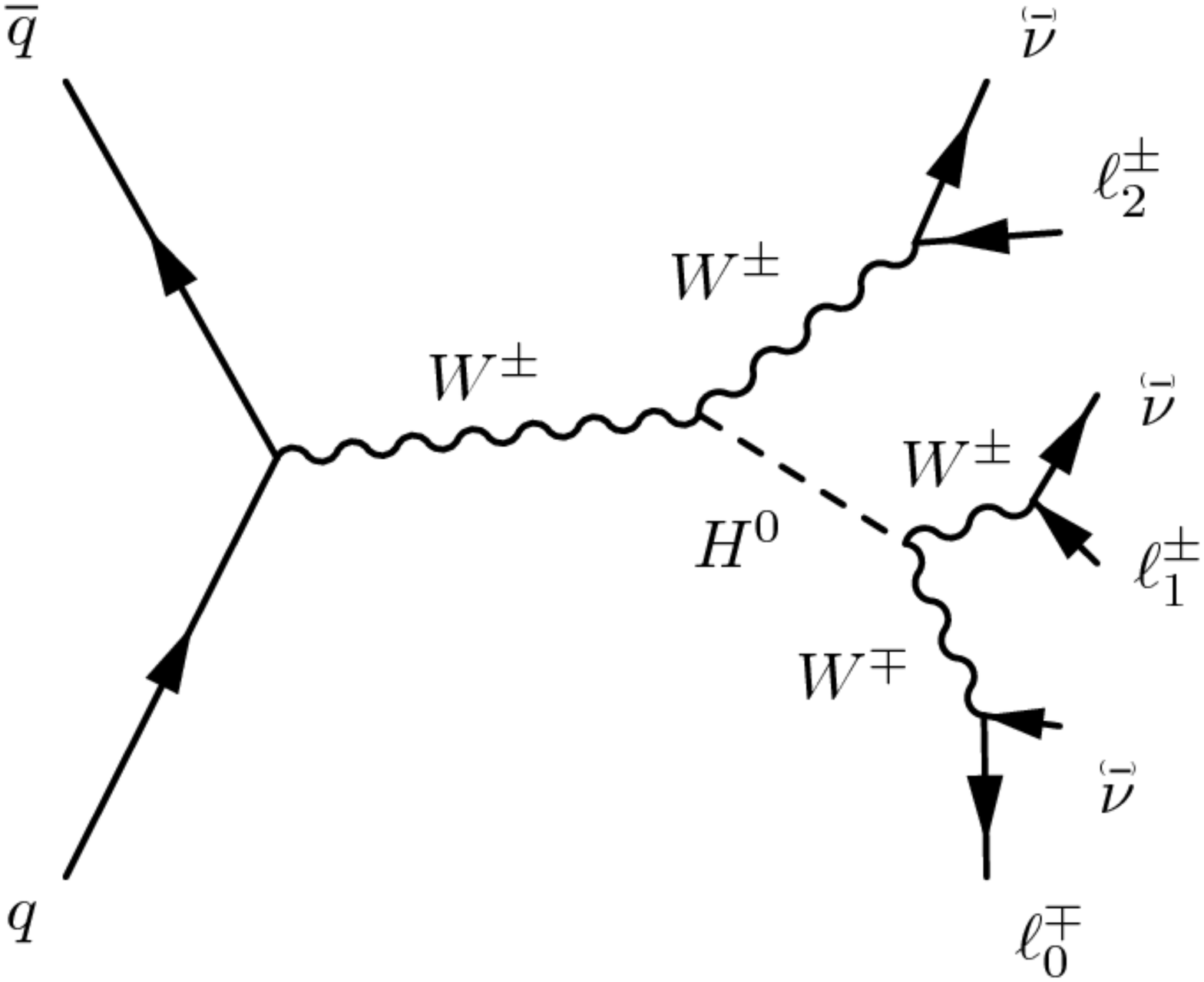}\hspace*{0.1\linewidth}
}
\subfloat[] {
\includegraphics[width=0.3\linewidth]{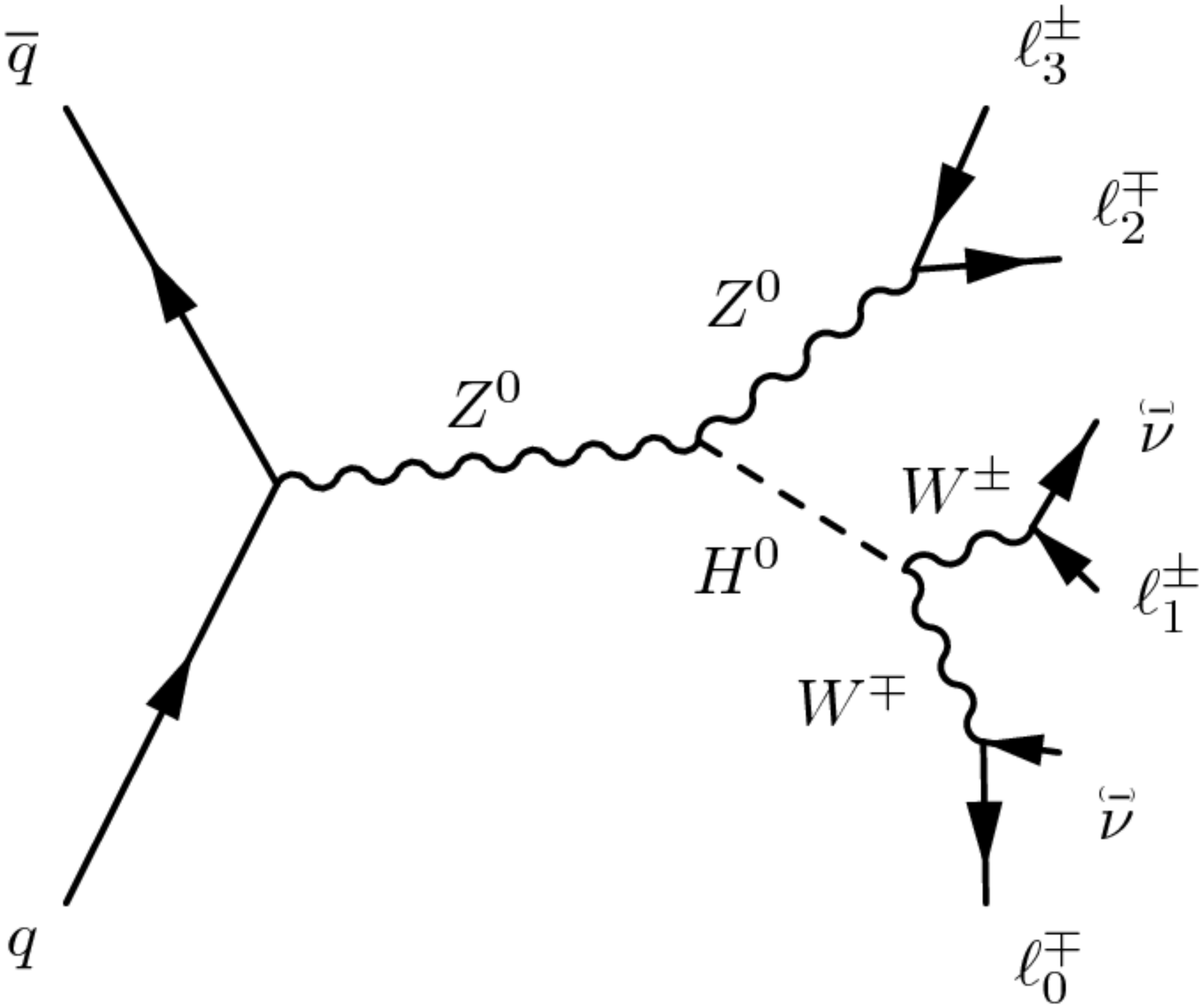}
}
\caption{\label{fig:feynman_diagrams}Tree-level Feynman diagrams
for the $VH(\HWW)$ topologies considered in this paper: (a) $3\ell$
channel and (b) $4\ell$ channel.}
\end{figure}

The analysis is performed using events with three ($3\ell$) or four
($4\ell$) charged leptons (electrons or muons) in the final state, targeting the \WH and \ZH
channels respectively.
Leptonic decays of $\tau$ leptons, from
\HWWtntn or \HWWtnln or from the associated vector bosons, are considered as signal, while no specific selection
is performed for events with hadronically decaying $\tau$ leptons in
the final state. Events from \VH production with \Htautau are considered as
background. The leading-order Feynman diagrams for the \WH and \ZH production
processes are depicted in Figure~\ref{fig:feynman_diagrams}.

In the \WH channel, multivariate discriminants are used to maximise
the sensitivity to the Higgs boson signal, while in the \ZH channel the analysis
is performed through selection requirements. The distribution of these \WH
discriminants, together with event counts in background control
regions and the signal regions in the \ZH channel, are combined
in a binned maximum-likelihood fit to extract the signal yield and
the background normalisations. The maximum-likelihood fit provides
results for the \WH and the \ZH channels separately and for their
combination \VH, assuming the Standard Model (SM) prediction for the relative cross sections of the two production processes.
\section{ATLAS detector}
\newcommand{\AtlasCoordFootnote}{
ATLAS uses a right-handed coordinate system with its origin at the nominal interaction point (IP)
in the centre of the detector and the $z$-axis along the beam pipe.
The $x$-axis points from the IP to the centre of the LHC ring,
and the $y$-axis points upwards.
Cylindrical coordinates $(r,\phi)$ are used in the transverse plane,
$\phi$ being the azimuthal angle around the $z$-axis.
The pseudorapidity is defined in terms of the polar angle $\theta$ as $\eta = -\ln \tan(\theta/2)$.
Angular distance is measured in units of $\Delta R \equiv \sqrt{(\Delta\eta)^{2} + (\Delta\phi)^{2}}$.}

The ATLAS experiment~\cite{PERF-2007-01,ATLAS-TDR-19, Abbott:2018ikt} is a multi-purpose particle detector
with a forward--backward symmetric cylindrical geometry and a near $4\pi$ coverage in
solid angle.\footnote{\AtlasCoordFootnote ~Transverse momentum and energy are defined as $\pT = p \sin \theta$ and $\et = E \sin \theta$ respectively.}
It consists of an inner tracking detector (ID) surrounded by a thin superconducting solenoid
providing a \SI{2}{\tesla} axial magnetic field, electromagnetic (EM) and hadronic calorimeters, and a muon spectrometer (MS).
The inner tracking detector covers the pseudorapidity range $|\eta| < 2.5$.
It consists of silicon pixel, silicon micro-strip, and transition-radiation tracking detectors.
Lead/liquid-argon (LAr) sampling calorimeters provide electromagnetic energy measurements
with high granularity.
A hadronic (steel/scintillator-tile) calorimeter covers the central pseudorapidity range ($|\eta| < 1.7$).
The endcap and forward regions are instrumented with LAr calorimeters
for both EM and hadronic energy measurements up to $|\eta| = 4.9$.
The muon spectrometer surrounds the calorimeters and is based on
three large air-core toroidal superconducting magnet systems that
provide a field integral  between \num{2.0} and \SI{6.0}{\tesla\metre}
across most of the detector.
The muon spectrometer includes a system of precision tracking chambers
covering the region $|\eta| < 2.7$ and fast detectors for triggering
within the range $|\eta| < 2.4$.
A two-level trigger system is used to select events~\cite{TRIG-2016-01}.

\section{Signal and background Monte Carlo simulation}
\label{sec:DataMC}
 
Monte Carlo (MC) event generators are used to model signal and background processes. All signal samples were generated with a Higgs boson mass of 125~\GeV~\cite{HIGG-2014-14,HIGG-2016-33}.
For most processes, separate programs were used to generate the hard scattering process and to model the underlying event and  the parton showering (UEPS). A description of the MC samples is given in Table~\ref{tab:mcsamples}. They are normalised to cross-section predictions calculated with the QCD and electroweak (EW) orders specified in the last column of Table~\ref{tab:mcsamples}.

\begin{table}
\centering
\caption{Monte Carlo generators used to model the signal and background processes. Alternative generators, underlying event and parton-showering  models, used to estimate systematic uncertainties, are shown in parentheses. In the last column the prediction order for the total cross section is shown. ``\PythiaSix'' refers to version 6.428, ``\PythiaEight'' refers to versions 8.210 or 8.186.
}
{\footnotesize
\begin{tabular}{llllc}
\hline\hline
Process & Generator  & UEPS model & Prediction order  \\
& (alternative) & (alternative) & for total cross section \\
\hline
 
$q\bar{q}\rightarrow WH$     & \textsc{Powheg-Box} v2 MiNLO       & \PythiaEight   & NNLO QCD + NLO EW~\cite{HanWillenbrock1991,BreinDjouadiHarlander2004,CiccoliniDittmaier2003} \\
&                                & (\textsc{Herwig 7})     &        \\
$q\bar{q}\rightarrow ZH$     & \textsc{Powheg-Box} v2 MiNLO       & \PythiaEight   & NNLO QCD + NLO EW~\cite{HanWillenbrock1991,BreinDjouadiHarlander2004,CiccoliniDittmaier2003} \\
&                                & (\textsc{Herwig 7})     &        \\
$gg\rightarrow ZH$           & \textsc{Powheg-Box} v2           & \PythiaEight   & NLO + NLL~\cite{Altenkamp:2012sx} \\
&                                & (\textsc{Herwig 7})     &        \\
ggF $H$  & \textsc{Powheg-Box} v2 NNLOPS & \PythiaEight   & NNNLO QCD + NLO EW~\cite{Anastasiou2016} \\
VBF $H$  & \textsc{Powheg-Box} v2        & \PythiaEight   & NNLO QCD + NLO EW~\cite{CiccoliniDennerDittmaier2007}  \\
\hline
\hline

$\ttbar$          & \textsc{Powheg-Box} v2 & \PythiaEight   &  NNLO+NNLL~\cite{Czakon:2013goa} \\
&                        & (\textsc{Herwig 7})     &             \\
& (\SHERPA 2.2.1)           & (\SHERPA 2.2.1)     &             \\
$Wt$              & \textsc{Powheg-Box} v1 & \PythiaSix     &  NLO~\cite{Re:2010bp} \\
$\ttbar W/Z$      & \MGFiveLO             & \PythiaEight   & NLO~\cite{Frixione:2015zaa,Alwall:2014hca}  \\
$tZ$              & \MGFiveLO             & \PythiaSix     & LO~\cite{Campbell:2013yla}  \\
$q\bar{q}/qg \to  \ell\nu\ell\ell$       & \SHERPA 2.2.2  & \SHERPA 2.2.2  & NLO~\cite{Cascioli:2013gfa}   \\
& (\textsc{Powheg-Box} v2) & (\textsc{Herwig++})      &       \\
$q\bar{q}/qg \to  \ell\ell\ell\ell$  & \SHERPA 2.1/2.2.2 & \SHERPA 2.1/2.2.2 & NLO~\cite{Cascioli:2013gfa}   \\
& (\textsc{Powheg-Box} v2) & (\textsc{Herwig++})     &       \\
$gg \to  \ell\ell\ell\ell$  & \SHERPA 2.1.1 & \SHERPA 2.1.1 & NLO~\cite{Caola:2016trd}  \\
 
$VVV$ & \SHERPA 2.2.2 & \SHERPA 2.2.2 & NLO~\cite{Binoth:2008kt}   \\
& (\MGFive) &  (\PythiaEight)   &       \\
 
\hline\hline
\end{tabular}
}
\label{tab:mcsamples}
\end{table}

The $q\bar{q} \rightarrow WH$ and  $q\bar{q} \rightarrow ZH$ processes were generated
with \textsc{Powheg-Box} v2~\cite{Luisoni:2013kna} MiNLO interfaced to \PythiaEight~\cite{pythia8}, with the AZNLO set of tuned parameters (tune)~\cite{STDM-2012-23} for underlying event, showering and hadronisation.
The $gg \to ZH$ process was  simulated with  \textsc{Powheg-Box} v2 + \PythiaEight with the AZNLO tune for underlying event, showering and hadronisation.
For the $VH$ samples, the PDF4LHC15 parton distribution function (PDF) set~\cite{PDF4LHC15} was used for the hard scattering process in  \textsc{Powheg-Box}~v2 and the  CTEQ6L1 PDF set~\cite{Pumplin:2002vw} was used for the parton showering in \PythiaEight.
\textsc{Herwig~7} ~\cite{Bellm:2015jjp}, with the MMHT2014lo68cl PDF set~\cite{Harland-Lang:2014zoa}, was used as an alternative parton-showering model for $VH$. The uncertainty due to the PDF choice is smaller than the uncertainty obtained by using \textsc{Herwig} as an alternative parton shower model (Section~\ref{sec:systematics}).
 
The gluon--gluon fusion (ggF) events were generated with \textsc{Powheg-Box} v2~NNLOPS~\cite{Hamilton:2013fea} interfaced to \PythiaEight  with the AZNLO tune.
The vector-boson fusion (VBF) events were generated with \textsc{Powheg-Box}~v2, interfaced to \PythiaEight.
For the ggF and VBF samples, the PDF4LHC15  PDF set was used for the hard scattering process in  \textsc{Powheg-Box} v2 and the  CTEQ6L1 PDF set was used for the parton showering in \PythiaEight. The contribution from the $\ttbar H$ and $tH$ production modes is negligible.

The top-quark pair production ($\ttbar$) was simulated with \textsc{Powheg-Box} v2~\cite{Frixione:2007nw} using the NNPDF 3.0 NNLO PDF set~\cite{Ball:2014uwa} and interfaced to \PythiaEight using the NNPDF 2.3 PDF set~\cite{Ball:2012cx} for parton showering, with the A14 tune~\cite{ATL-PHYS-PUB-2014-021}.
For $\ttbar$ production, \SHERPA~\cite{Gleisberg:2008ta} 2.2.1, with the NNPDF 3.0 PDF set, was used as an alternative generator while \textsc{Herwig 7}, with the MMHT2014lo68cl PDF set, was used as an alternative UEPS model.
The single-top-quark production  $Wt$ was generated with
\textsc{Powheg-Box} v1~\cite{Re:2010bp} interfaced to  \PythiaSix~\cite{Sjostrand:2006za} for parton showering with the \textsc{Perugia}2012 tune~\cite{Skands:2010ak}.
\textsc{EvtGen}\,1.2.0~\cite{Lange:2001uf} was used for the simulation of $\textit{b}$-quark and $\textit{c}$-quark decays.
The $\ttbar W/Z$ and $tZ$ processes were generated at leading order (LO) with \MGFiveLO~\cite{Alwall:2014hca}~version 2.2.2
($\ttbar W/Z$) and 2.2.1 ($tZ$)
interfaced to \PythiaEight ($\ttbar W/Z$) and \PythiaSix ($tZ$), using the NNPDF2.3 LO PDF set.

The $q\bar{q}/qg \to VV^{\ast}$  samples with final states   $\ell\nu\ell\ell$ and $\ell\ell\ell\ell$~\cite{ATL-PHYS-PUB-2016-002} were generated with \SHERPA 2.2.2, with the exception of the $ZZ^{\ast}$ sample in the $WH$ analysis for which \SHERPA 2.1 was used; the CT10 PDF set~\cite{Lai:2010vv} and the NNPDF 3.0 PDF set were used for versions 2.1 and 2.2.2, respectively.
\textsc{Powheg-Box} v2~\cite{Nason:2013ydw} was used as an alternative generator for $VV^{\ast}$, with \textsc{Herwig++}, using the CTEQ6L1 PDF set, for parton showering. Among the loop-induced $gg$-initiated diboson processes, the only relevant  process in this analysis is $gg \to ZZ^{\ast}$, for which a $K$-factor of 1.55 was used~\cite{Caola:2016trd}. This process was simulated with \SHERPA 2.1.1, using the CT10 PDF set.

The triboson $VVV$ samples were  generated with \SHERPA 2.2.2 and the NNPDF 3.0 PDF set. \MGFive\ was used as an alternative generator
for $VVV$, with \PythiaEight, using the NNPDF2.3 LO PDF set.
The same PDF sets were used for the hard scattering and the parton showering in all the \SHERPA samples described above.

All simulated samples include the effect of pile-up from multiple interactions in the same and neighbouring bunch crossings.
This was achieved by overlaying minimum-bias events, simulated using \PythiaEight with the A2 tune~\cite{ATL-PHYS-PUB-2012-003} and MSTW2008LO PDF set~\cite{Martin:2009iq}. All samples were processed through the \GEANT4~\cite{Agostinelli:2002hh}  ATLAS detector simulation~\cite{SOFT-2010-01}.

\section{Event reconstruction}
 
Candidate signal events are selected using triggers that require a single isolated lepton with minimum transverse momentum (\pt) thresholds between 24~\gev\ and 26~\gev\ for electrons and between 20~\gev\ and 26~\gev\
for muons, depending on the data-taking period.
At least one of the leptons reconstructed offline is required to have triggered  the event and to have a \pt higher than the nominal trigger threshold by at least 1~\gev.
The single-lepton trigger efficiencies on the plateau are approximately 70\% for single muons with
$|\eta|<1.05$, 90\% for single muons in the range $1.05<|\eta|< 2.40$ and greater than 90\% for single electrons in the range $|\eta|<2.47$. The trigger efficiency for the signal events, estimated after the preselection, is 94\% for $WH$ and 98.5\% for $ZH$.
 
Selected events are required to have at least one primary vertex reconstructed from at least two associated tracks,
each with transverse momentum $\pt>400~\MeV$, as described in Ref.~\cite{Aaboud:2016rmg}.
If an event has more than one reconstructed primary vertex,
the vertex with the largest track $\sum \pt^2$ is selected for the analysis.

Electrons are reconstructed from clusters of energy deposits in the EM calorimeter matched to ID tracks,
and are identified using criteria based on the calorimeter shower shape, the quality of the match
between the track and the cluster and the amount of transition radiation emitted in the ID, as described in Ref.~\cite{PERF-2017-01}.
Electrons are required to satisfy $|\eta|<2.47$, excluding $1.37<|\eta|<1.52$,
which corresponds to the transition region between the barrel and the endcap EM calorimeters.
The efficiency for electron identification ranges from 88\% to  94\%, depending on electron  $\pt$ and $\eta$.
Muons are reconstructed by combining ID and MS tracks with consistent trajectories and curvatures.
An overall fit of hits from the ID track, energy loss in the calorimeter
and the hits of the track in the muon system is used to form muon candidates, as described in Ref.~\cite{PERF-2015-10}.  The efficiency for muon identification is close to  95\% over the full instrumented  $\eta$ range.
To suppress particles misidentified as leptons, several identification requirements as well as impact parameter, calorimeter and track isolation criteria~\cite{PERF-2017-01,PERF-2015-10} are applied.
 
Jets are reconstructed using the anti-$k_{t}$ algorithm with radius parameter $R=0.4$~\cite{Cacciari:2008gp,Cacciari:2011ma}.
The four-momenta of jets are corrected for the effects of calorimeter non-compensation,
energy loss in non-instrumented regions, and contributions from pile-up~\cite{PERF-2016-04}.
Jets are required to have $|\eta| < 4.5$, with $\pt > 25$~\GeV\ for the region  $|\eta|<2.5$ and  $\pt > 30$~\GeV\ for the region $2.5<|\eta|<4.5$.
A multivariate selection~\cite{PERF-2014-03} is used to suppress jets with $\pt < 60$~\GeV\ and $|\eta| < 2.4$ originated from pile-up.
Furthermore, to suppress pile-up jets in the forward region, jet shapes and topological jet correlations in
pile-up interactions are exploited~\cite{PERF-2016-06-witherratum}. Jets with \pt > 20~\gev\  and  $|\eta| < 2.5$ containing $b$-hadrons ($b$-jets) are identified using a multivariate technique~\cite{PERF-2016-05} with an efficiency of 85\%, estimated from simulated $t\bar{t}$ events. The multivariate technique gives rejection factors 
against jets originating from a light quark or gluon and jets containing $c$-hadrons of 33 and 3, respectively. 
 
The missing transverse momentum \pmet with magnitude  \vmet\ in each event is calculated from
the negative vectorial sum of the transverse momenta of electrons, muons, and jets. It uses both track-based and calorimeter-based measurements~\cite{PERF-2016-07}.

\section{Event selection}
\label{sec:analysis}
In the \WH channel, exactly three isolated leptons with  $\pt > 15~\gev$  are required with a total charge of $\pm1$. The lepton with unique charge is labelled $\ell_0$, the lepton closest to $\ell_0$ in angular distance $\Delta R$ is labelled $\ell_1$, and the remaining lepton is labelled $\ell_2$. In signal events leptons $\ell_0$ and $\ell_1$ are most likely to originate from the $\hww$ decay, with probabilities of $99\%$ and $85\%$ respectively.
 
The most prominent background processes to the \WH channel are $WZ/W\gamma^{\ast}$ production and top-quark processes with either three prompt leptons, e.g.\ $t\bar{t}V$, or two prompt leptons and one non-prompt lepton from a $b$-hadron decay, e.g.\ $t\bar{t}$. Other important background processes  are $ZZ^{\ast}$ (including $Z\gamma^{\ast}$), $Z\gamma$ and $Z$+jets production; they may satisfy the signal selection requirements if a lepton is undetected, in the case of $ZZ^{\ast}$, or if they contain a misidentified or non-prompt lepton, in the case of $Z\gamma$ and $Z$+jets production. Processes with three prompt leptons in the final state such as tribosons, in particular $WWW$,  also contribute to the background.
Contributions from background processes that include more than one misidentified lepton, such as  $W$+jets production and inclusive $b{\bar b}$ pair production, are negligible.
The background from top-quark production is suppressed by vetoing events if they contain any $b$-tagged jet.
 
\begin{figure}
\centering
\subfloat[] {
\includegraphics[width=0.49\linewidth]{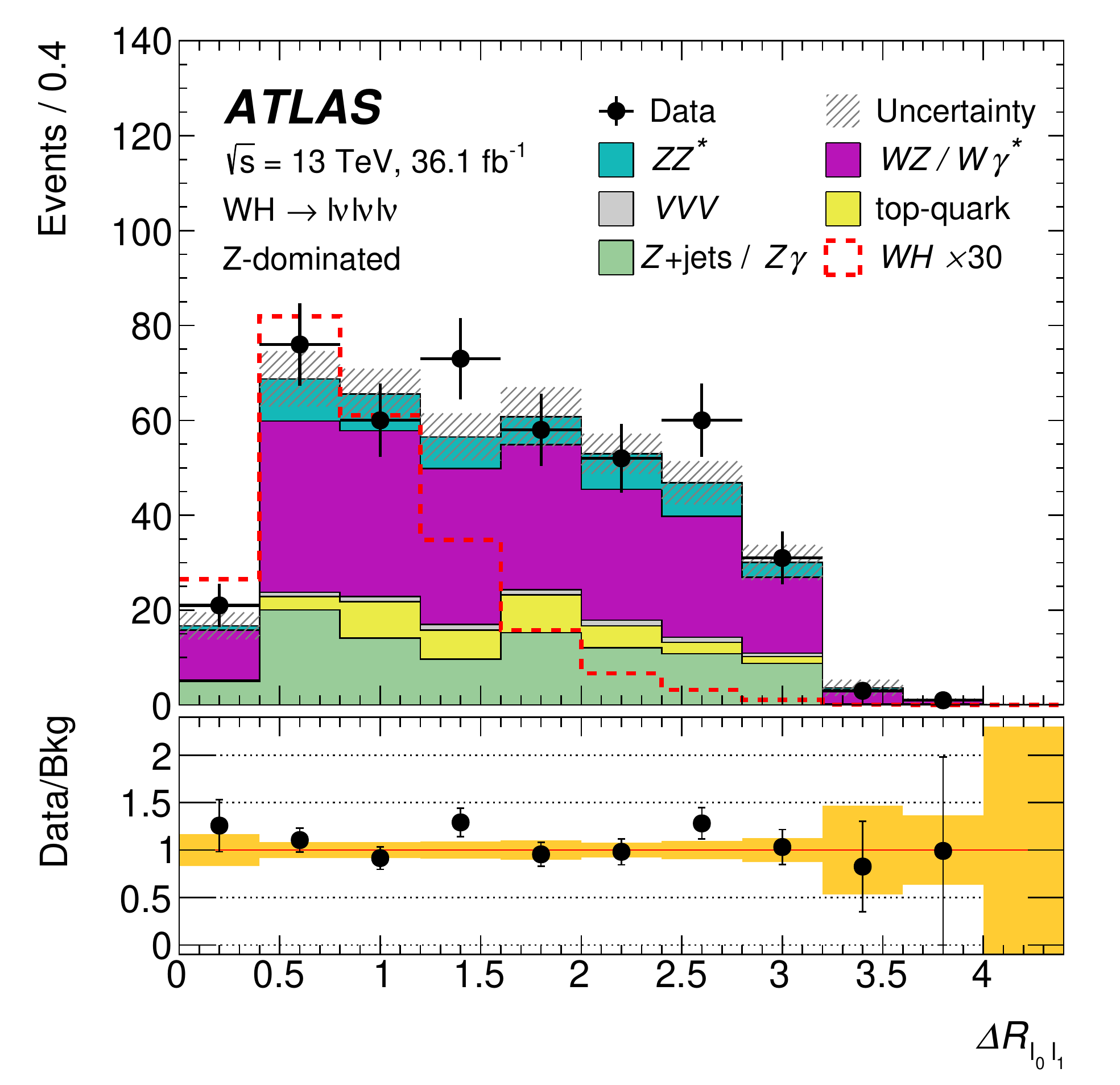}
}
\subfloat[] {
\includegraphics[width=0.49\linewidth]{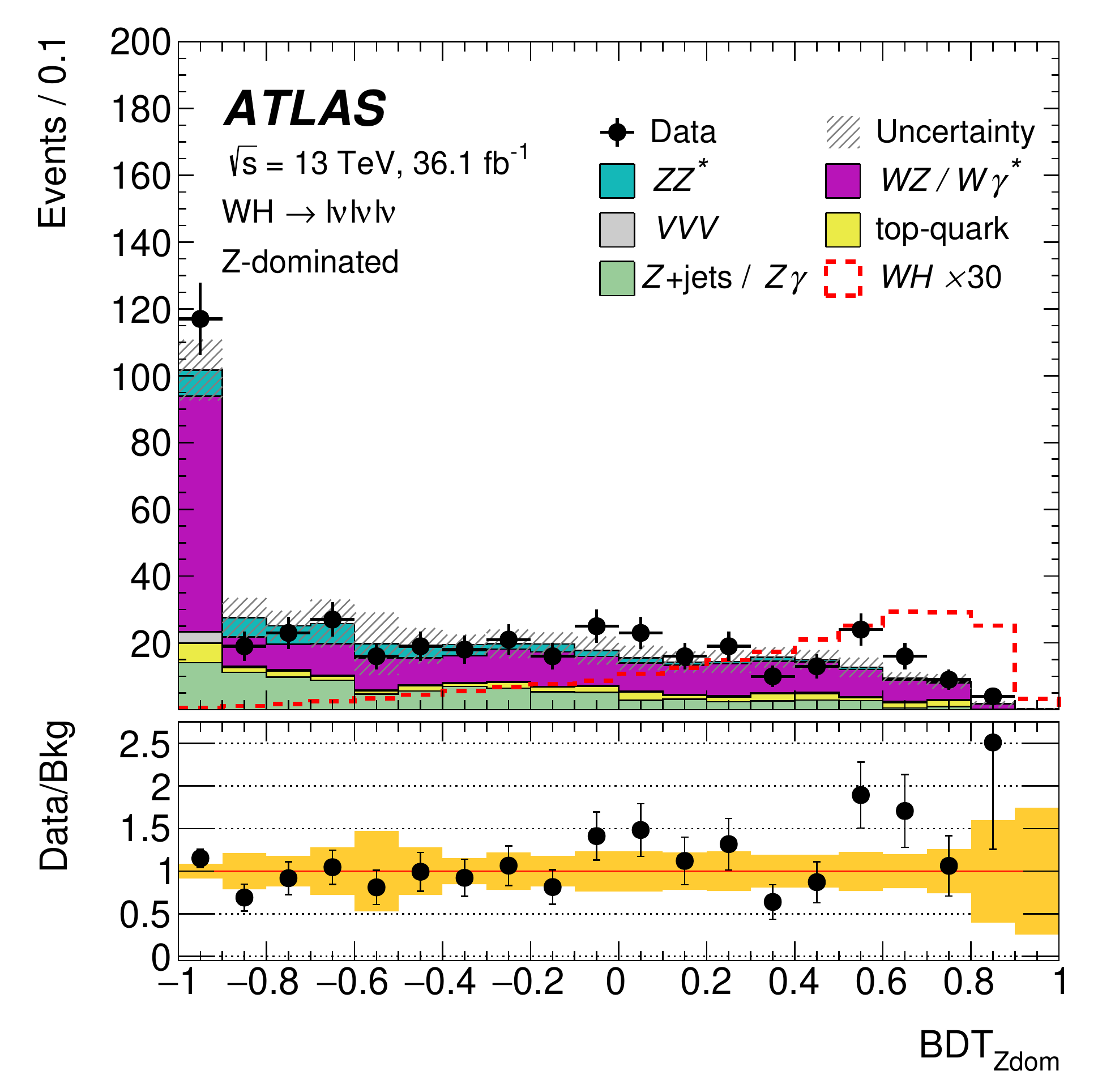}
}
\caption{\label{fig:bdtSR12} Distribution of (a) angular separation $\Delta R_{\ell_0\ell_1}$ and (b) BDT$_\textrm{Zdom}$  distribution in the $Z$-dominated category. The dashed line shows the \WH signal scaled by a factor of 30.
The hatched band in the upper panel and the shaded band in the lower panel show the total statistical and experimental systematic uncertainty in background predictions.
The $WZ/W\gamma^{\ast}$ and top-quark background processes are normalised with the normalisation factors from the control region analysis (Table ~\ref{tab:CRsSelection}).
The $Z$+jets and $Z\gamma$ background processes are estimated with the data-driven technique described in Section~\ref{sec:background}. }
\end{figure}

The analysis of the \WH channel separates events with at least one same-flavour opposite-sign charge (SFOS) lepton pair from events with zero SFOS lepton pairs, which have different signal-to-background ratios.
Due to the presence of $Z\rightarrow\ell\ell$ decays as a dominant background, the former is hereafter referred to as the $Z$-dominated category, while the latter is referred to as the $Z$-depleted category.
In the $Z$-dominated category, the major background processes are those involving $Z$ bosons. Events are vetoed if they contain more than one jet. This requirement further suppresses top-quark events  with an additional non-prompt lepton from $b$-hadron decays. In order to select final states with neutrinos, \met{} is required to be above 30~\gev. The invariant masses $m_{\ell\ell}$ of all SFOS pairs are required to satisfy  a $Z$-veto selection: $|m_{\ell\ell}- m_{Z}| > 25~\gev$. The last two requirements suppress $WZ/W\gamma^{\ast}$ and $ZZ^{\ast}$ events, and improve the $Z$+jets rejection.
In order to suppress background events from heavy-flavour quarkonia, the smallest invariant mass of SFOS pairs is required to be greater than 12~\gev. The signal efficiency of this selection with respect to the preselection is $34.6\%$.
A discriminant based on a boosted decision tree (BDT)~\cite{Hoecker:1019880,scikit-learn} is used to achieve a further separation between signal and background processes.
The main purpose of the multivariate classifier, named BDT$_\textrm{Zdom}$, is to distinguish between signal and the dominant $WZ/W\gamma^{\ast}$  and $ZZ^{\ast}$ background processes, and hence it is trained against these two background processes.
The BDT$_\textrm{Zdom}$ uses seven input variables. They
are the magnitude of the vectorial sum of lepton transverse momenta ($|\Sigma{\pt^{\ell_i}}|$), the invariant masses of the first lepton pair ($m_{\ell_0\ell_1}$) and of the three leptons  ($m_{\ell\ell\ell}$), the angular distance  $\Delta R_{\ell_0\ell_1}$, $\met$ and the pseudorapidity separation between the leptons with the same charge ($\Delta \eta_{\ell_1\ell_2}$). Moreover the BDT uses the transverse mass of the $W$ boson, $m_{\textrm{T}}^W=\sqrt{2\,\pt^{\ell_W}\cdot\met\cdot\left(1-\cos{\Delta\phi\left({\ell_W},\pmet\right)} \right)}$, built from the $\pmet$ and the lepton $\ell_W$ which is the lepton not belonging to  the SFOS lepton pair with invariant mass closer to the $Z$ boson mass, and could be either $\ell_1$ or $\ell_2$. Figure~\ref{fig:bdtSR12} shows the distribution of $\Delta R_{\ell_0\ell_1}$, which is the most powerful  variable in the BDT$_\textrm{Zdom}$ training,   and  the BDT$_\textrm{Zdom}$ distribution in the $Z$-dominated category, before applying any selection requirement on the BDT$_\textrm{Zdom}$ score.
 
The $Z$-dominated signal region (SR), defined as the events with high-ranking BDT$_\textrm{Zdom}$ score (BDT$_\textrm{Zdom}$$>0.3$), is divided into three bins with increasing sensitivity: $0.3\leq \mathrm{BDT_\textrm{Zdom}} <0.5$,  $0.5\leq \mathrm{BDT_\textrm{Zdom}} <0.7$ and $0.7\leq \mathrm{BDT_\textrm{Zdom}} <1.0$. This binning is the result of an optimisation to find the minimal number of BDT bins that gives the highest sensitivity.
The expected signal-to-background ratio in these bins is about 0.05, 0.09 and 0.19, respectively.
The full $Z$-dominated event selection is summarised in Table~\ref{tab:SRsSelection}.
 
\begin{table}
\begin{center}
\caption{ Event selection criteria used to define the signal regions in the \WH and \ZH analyses. The symbols are defined in the text.}
\resizebox{\textwidth}{!}{
\begin{tabular}{l||cccc}
\hline \hline
& \multicolumn{2}{c}{ $WH$} & \multicolumn{2}{c}{ $ZH$ } \\
 
\hline
Preselection & \multicolumn{2}{c}{3 isolated leptons (\pt$>$ 15 ~\gev)}  & \multicolumn{2}{c}{4 isolated leptons (\pt$>$ 10 ~\gev)} \\
& \multicolumn{2}{c}{total lepton charge $\pm1$}  & \multicolumn{2}{c}{total lepton charge 0} \\

\hline
Category                          & $Z$-dominated     &  $Z$-depleted    &  2-SFOS &   1-SFOS     \\
\hline
Number of SFOS                &2 or 1                        &0                   & 2         &   1         \\
Number of jets                &$\le1$                    & ---                  &$\le1$     &  $\le2$   \\
Number of $b$-jets             &0                        &0                &0                        &0         \\
\met\ [\gev]                  & $>30$                    &  ---                          &$>45$ &   ---     \\
$p_{\textrm{T}}^{4\ell}$ [\gev]     & --- & --- & $>45$ & ---\\
$m_{\ell\ell}$ [\gev]         & $>12$ (min. SFOS)           & ---   & $>10$   & $>10$     \\
$|m_{\ell\ell} - m_{Z}|$ [\gev] &  $>25$ (SFOS)           & ---            &    $<10~(m_{\ell_2\ell_3})$ & $<10~(m_{\ell_2\ell_3})$ \\
$m_{\ell_0\ell_1}$ [\gev]        & ---                           &  ---        &   $<55$ &  $<60$   \\
 
$\Delta \phi_{\ell_0\ell_1}^{\mathrm{boost}}$ & --- & --- & $<2.3$ & $<1.9$\\
$m_{\tau\tau}$ [\gev]     & --- & --- & --- & $<50$\\
$\Delta\phi_{\ell_0\ell_1,\,\pmet}$ [rad]       & --- & --- & --- & $>0.4$\\
$m_{4\ell}$ [\gev]     & --- & --- & $>140$ & ---\\
BDT                & BDT$_\textrm{Zdom}>0.3$ & $\bdttop>0.2$ $\&$ $\bdtwz>0.15$  & --- & ---\\
\hline\hline
\end{tabular}
}
\label{tab:SRsSelection}
\end{center}
\end{table}
 
 
In the $Z$-depleted category, the two major background processes are $WZ/W\gamma^{\ast}$ with $Z/\gamma^{\ast}\to\tau\bar{\tau}$ and \ttbar, where $WZ/W\gamma^{\ast}$ has the same signature of the signal, namely three prompt leptons, while \ttbar contains a misidentified or non-prompt lepton from a jet. Two separate BDTs, named \bdtwz and \bdttop, are trained 
to allow an optimal background rejection. The \bdtwz uses 11 input variables, of which three common to the BDT$_\textrm{Zdom}$ are $m_{\ell_0\ell_1}$, \met and $\Delta\eta_{\ell_1\ell_2}$; the other variables are
the transverse momenta of the three leptons ($\pt^{\ell_0}$, $\pt^{\ell_1}$, $\pt^{\ell_2}$), the transverse mass ($m^{\textrm{T}}_{\ell_0\ell1}$) built from $\ell_0$, $\ell_1$ and \pmet,
the invariant mass of the electrons with same-sign charge ($m_{ee}$),
the transverse impact parameter significances of the lepton with lowest $\pt$ ($|d_{\textrm{0,sig,min}}|$), the transverse impact parameter significances of the lepton with second-lowest $\pt$  and opposite charge with respect to the lepton with lowest $\pt$ ($|d_{\textrm{0,sig,mid}}|$) and the compatibility of the event with the $WZ$ hypothesis $F_{\alpha}$.\footnote{Given the reconstructed charged-lepton momenta and the \pmet, the event kinematics can be calculated under the $WZ$ with $Z\to\tau\tau$ hypothesis and using the collinear approximation for the $\tau$ decays with one remaining unknown, e.g.\ the ratio of one $\tau$ energy to the energy of the lepton from this $\tau$ decay. This unknown is varied and the
number of physical kinematic solutions is taken as a measure of the compatibility with the $WZ$ hypothesis.}

\begin{figure}
\centering
\subfloat[] {
\includegraphics[width=0.49\linewidth]{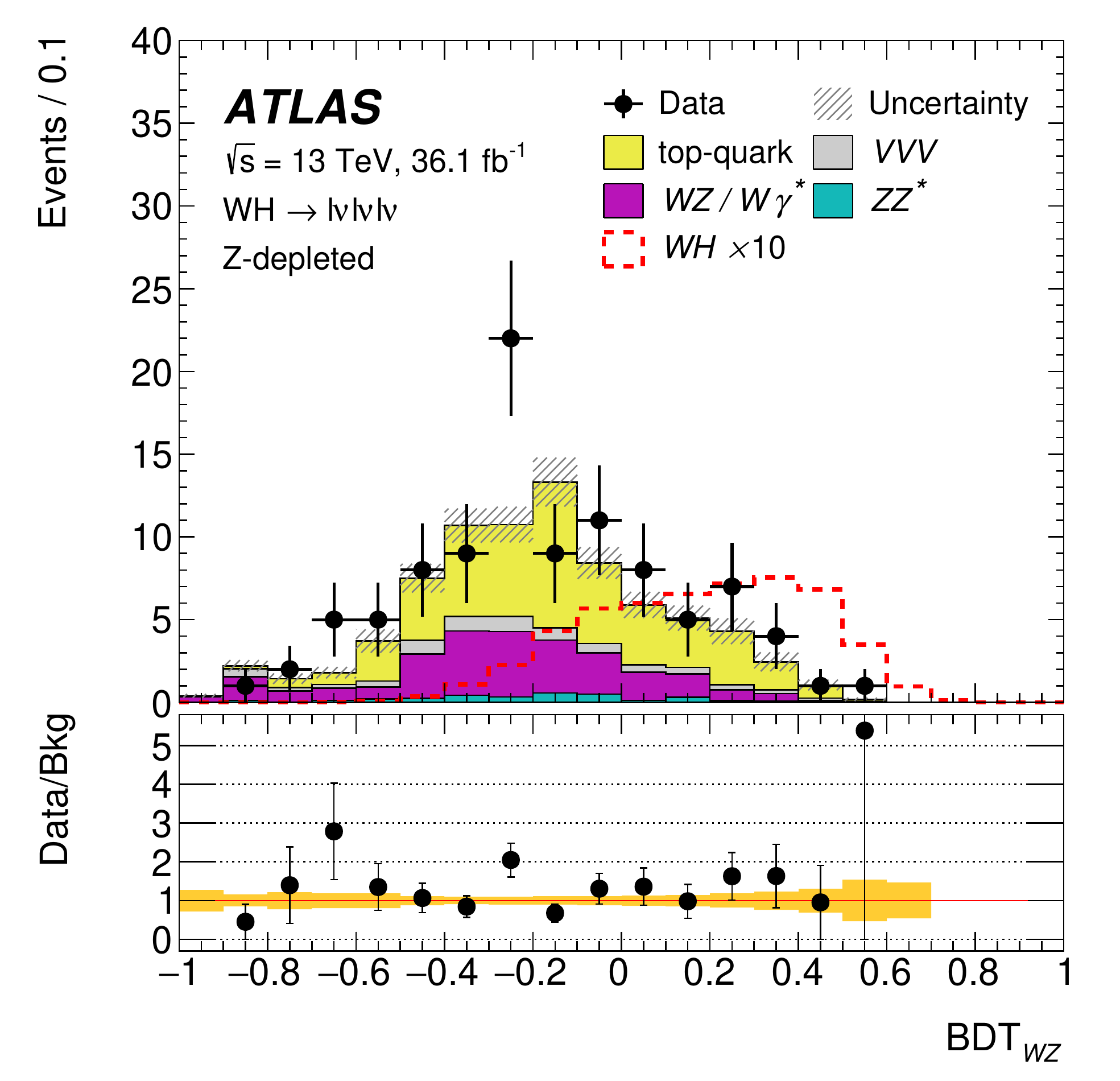}
}
\subfloat[] {
\includegraphics[width=0.49\linewidth]{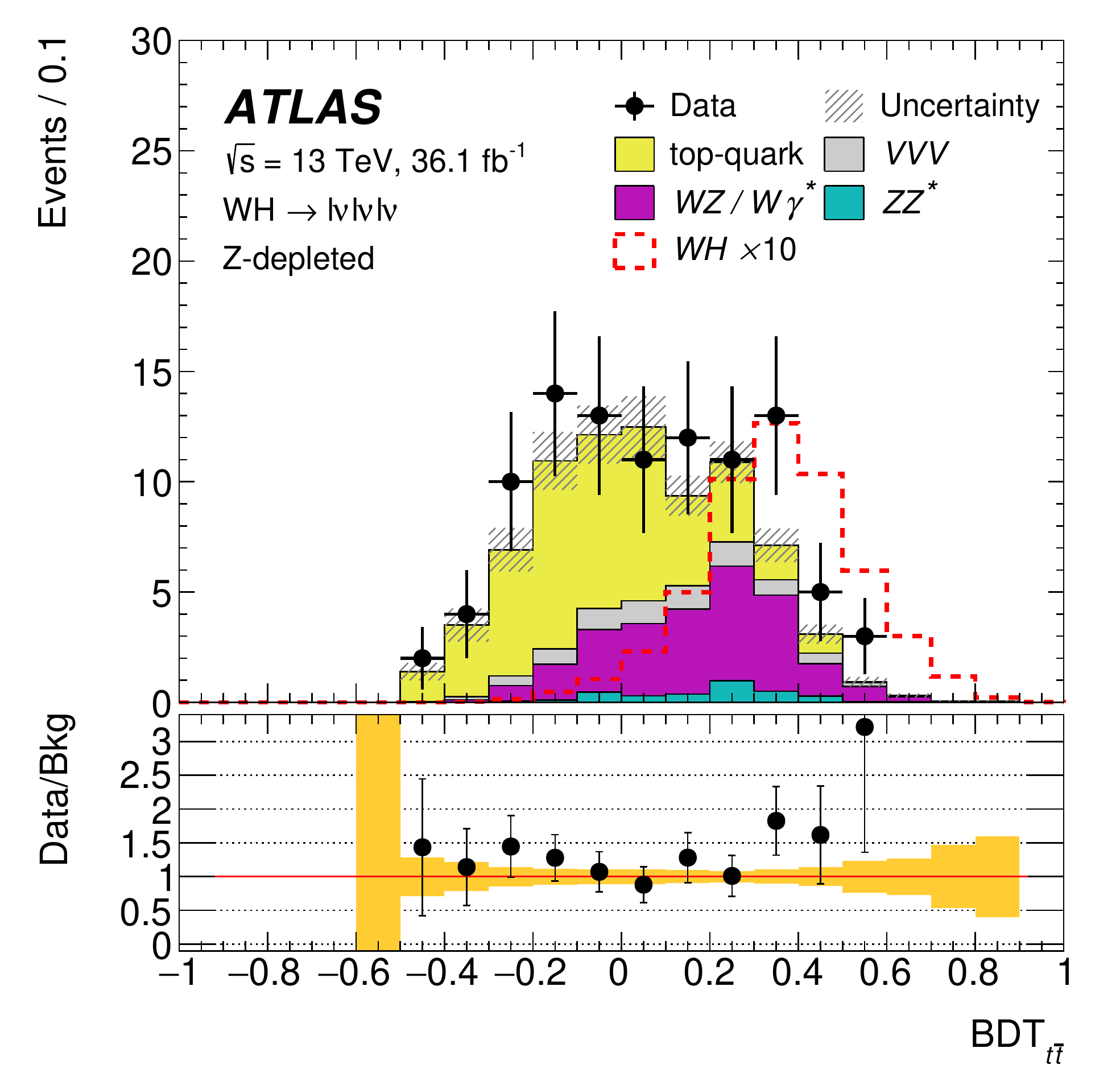}
}
\caption{\label{fig:bdtSR3}Distribution of (a) \bdtwz and (b) \bdttop in the $Z$-depleted category. The dashed line shows the \WH signal scaled by a factor of 10. The hatched band in the upper panel and the shaded band in the lower panel show the total statistical and experimental systematic uncertainty in background predictions.
The $WZ/W\gamma^{\ast}$ and top-quark background processes are normalised with the normalisation factors from the control region analysis (Table ~\ref{tab:CRsSelection}).}
\end{figure}

A definition of the most likely lepton from heavy-flavour decays ($\ell_{\textrm{HFL}}$) is crucial for an optimal performance of the \bdttop.
For this purpose, a \bdthfl is trained purely on data using track and calorimeter isolation as well as impact-parameter variables as input. The lepton with the minimal \bdthfl output is selected as $\ell_{\textrm{HFL}}$.
The \bdttop uses nine input variables, of which two common to the BDT$_\textrm{Zdom}$ and \bdtwz are $m_{\ell_0\ell_1}$ and $\Delta \eta_{\ell_1\ell_2}$, one common to the BDT$_\textrm{Zdom}$ is $\Delta R_{\ell_0\ell_1}$, and the other input variables are
the number of jets ($N_{\textrm{jet}}$), the transverse momentum of the leading jet ($\pt^{j_{\textrm{lead}}}$), the invariant mass of the leptons with same-sign charge ($m_{\ell_1\ell_2}$), and
three $\ell_{\textrm{HFL}}$-related variables: its \bdthfl output, its transverse momentum, $\pt^{\ell_{\textrm{HFL}}}$, and the invariant mass built from it together with the closest opposite-charge lepton ($m_{\ell_{\textrm{HFL}}\ell_{\textrm{cloc}}}$).
Figure~\ref{fig:bdtSR3} shows the outputs of \bdtwz and \bdttop in the $Z$-depleted category, before applying any selection requirements on the BDT scores.
 
The choice of input variables for the different BDTs was the result of an optimisation study where several thousand different BDTs using different set of input variables have been compared and the best performing BDTs have been selected for the final analysis.
 
The full $Z$-depleted event selection is summarised in Table~\ref{tab:SRsSelection}. The events with high-ranking BDT scores (\bdttop> 0.2 and \bdtwz> 0.15) are used to define the $Z$-depleted SR. In this region, the BDT scores  are used as discriminant variables in the statistical analysis, with three bins in \bdttop, each being subdivided into two bins in \bdtwz  as shown in Table~\ref{tab:ZdeplBDTbins}.
The expected signal-to-background ratio in these bins ranges from about $0.07$ in the first bin up to about $3.6$ in the last bin. The signal efficiency with respect to the preselection, before any cut on the BDT scores,  is $23.5\%$.

The $ZH$ channel requires events to contain four isolated leptons with $p_{\textrm{T}} > 10$~\GeV\ and total electric charge of zero.
Events that contain a SFOS lepton pair with $m_{\ell\ell} < 10$~\GeV\ are rejected to suppress the contamination from heavy-flavour quarkonia.
Following this preselection, events are classified according to the number of SFOS lepton pairs: 1-SFOS and 2-SFOS. 
Events with no SFOS lepton pairs are not considered. 
 
The reconstruction of the $ZH$ process proceeds through the identification of the leptons from the $Z$ boson, called $\ell_2$ and $\ell_3$, as the SFOS lepton pair with invariant mass closest to the $Z$ boson mass, $m_Z$.
Then, the remaining two leptons, labelled  $\ell_0$ and $\ell_1$, are candidates for originating from the Higgs boson decay.
The background in the $ZH$ channel is almost exclusively due to $ZZ^{\ast}$ production. This constitutes $\sim$92\% of the total background after the preselection is applied. Processes with four prompt leptons in the final state such as triboson production, in particular $ZWW$, which has the same signature as the signal, and $t\bar{t}Z$ also contribute to the background. Other background processes such as $WZ/W\gamma^{\ast}$, $Z$+jets and $tV$ may contribute when at least one jet, hadron or a converted photon is misidentified as a lepton.

\begin{figure}
\centering
\subfloat[]{
\includegraphics[width=0.49\linewidth]{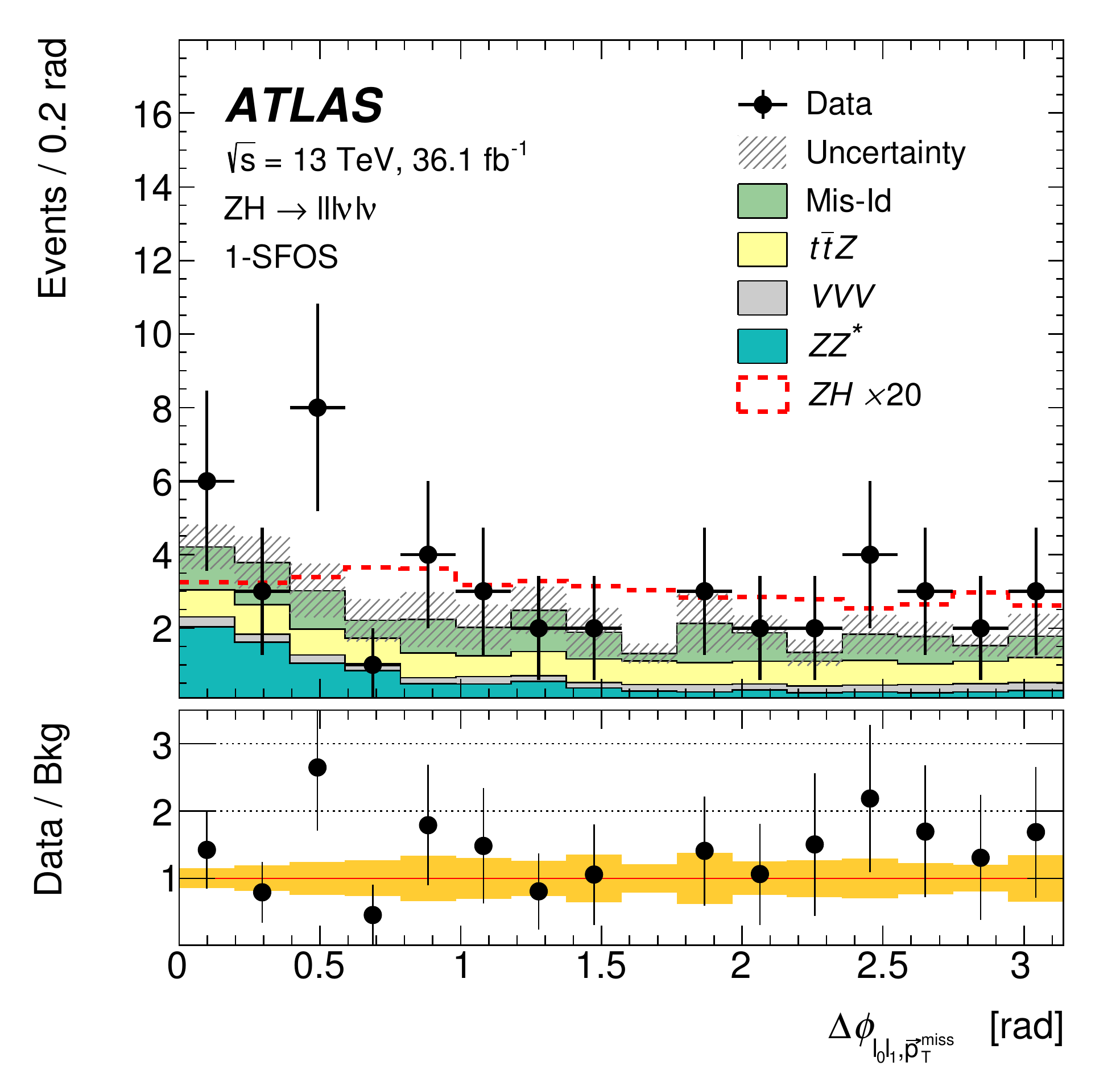}\label{fig:DPhillMET_1SFOS}
}
\subfloat[]{
\includegraphics[width=0.49\linewidth]{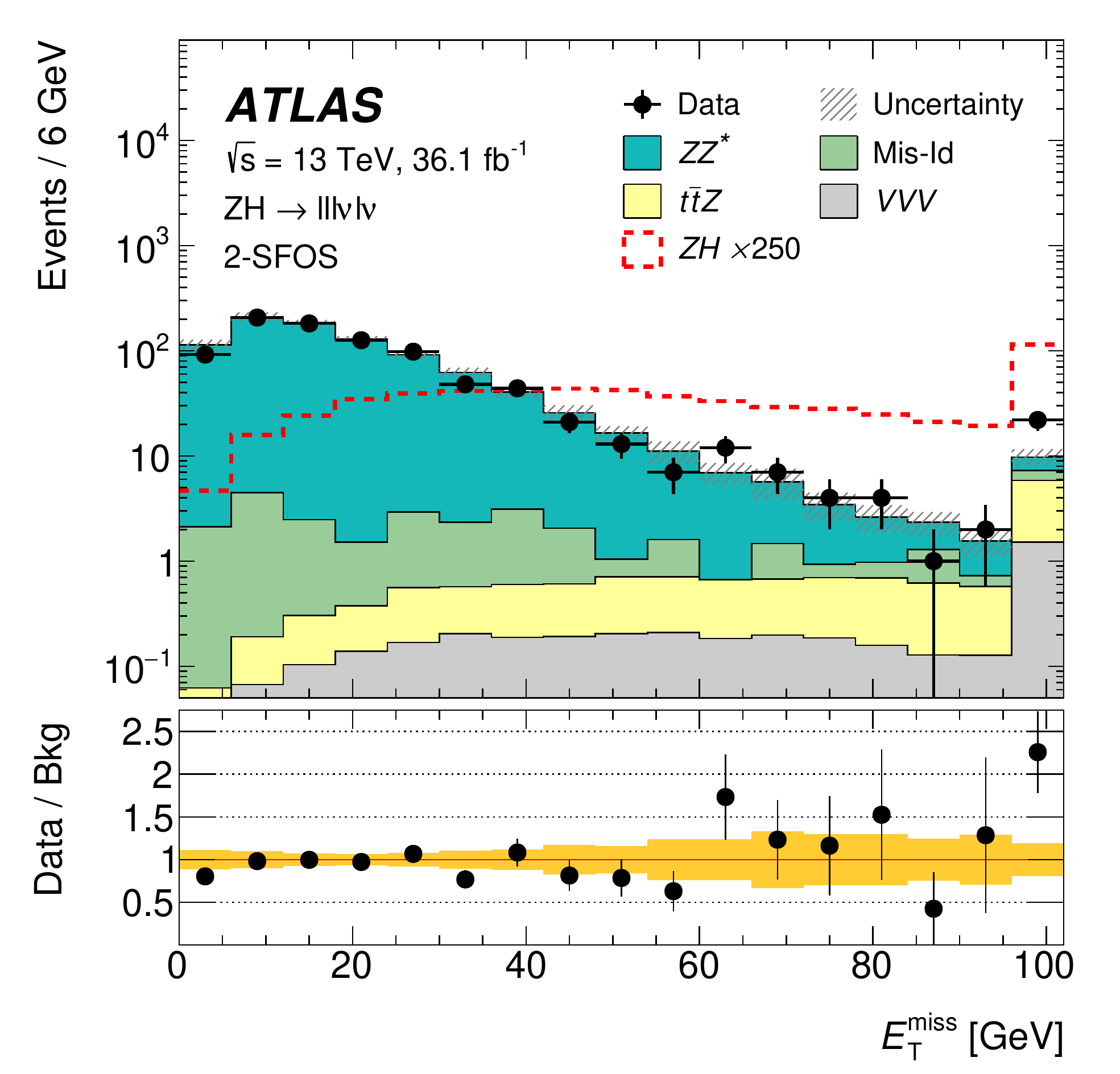}
\label{fig:MET_2SFOS}
}
\caption{Distributions of \protect\subref{fig:DPhillMET_1SFOS} the azimuthal separation between the Higgs-candidate lepton pair and the missing transverse momentum $\Delta\phi_{\ell_0\ell_1,\,\pmet}$ in 1-SFOS events and \protect\subref{fig:MET_2SFOS} the missing transverse momentum $E_{\textrm{T}}^{\textrm{miss}}$ in 2-SFOS events, after preselection. For the latter the last bin contains overflow events.
The dashed line shows the $ZH$ signal scaled by a factor 20 (left) and by a factor 250 (right). The hatched band in the upper panel and the shaded band in the lower panel show the total statistical and experimental systematic uncertainty in background predictions. The $ZZ^{\ast}$ background process is normalised in the 2-SFOS case with the normalisation factor from the control region analysis (Table~\ref{tab:CRsSelection}). The Mis-Id background is estimated with the data-driven technique described in Section~\ref{sec:background}. } 
\end{figure}

In order to suppress the $t\bar{t}Z$ process, events containing $b$-tagged jets are rejected and at most one and two jets are allowed in 2-SFOS and 1-SFOS classes, respectively. To reduce the $ZZ^{\ast}$ background process in events with two SFOS lepton pairs, a threshold of 45~\GeV\ is applied to the $E_{\textrm{T}}^{\textrm{miss}}$ and to the vector sum of the lepton transverse momenta, $p_{\textrm{T}}^{4\ell}$.  The invariant mass of $\ell_2$ and $\ell_3$, $m_{\ell_2\ell_3}$, is required to satisfy $|m_{\ell_2\ell_3} - m_Z| < 10$~\GeV, and the invariant mass of $\ell_0$ and $\ell_1$, $m_{\ell_0\ell_1}$, is required to be between 10~\GeV\ and 60~\GeV\ (55~\GeV) in 1-SFOS (2-SFOS) events. 
The variable $\Delta \phi_{\ell_0\ell_1}^{\mathrm{boost}}$ denotes the difference in azimuthal angle between the leptons from the Higgs boson candidate in the frame where the Higgs boson $p_{\textrm{T}}$ is zero. The Higgs boson transverse momentum is approximated with $\vec{p}_{\mathrm{T}}^{H} \approx - \vec{p}_{\mathrm{T}}^{Z}$, or $\vec{p}_{\mathrm{T}}^{H} \approx - \vec{p}_{\mathrm{T}}^{Z} - \sum \vec{p}_{\mathrm{T}}^{\mathrm{jet}}$ if at least one jet is present in the event.
Events are required to satisfy $\Delta \phi_{\ell_0\ell_1}^{\mathrm{boost}} < 1.9~(2.3)$~rad in the 1-SFOS (2-SFOS) class. 
In 1-SFOS events the $ZZ^{\ast}$ process contributes through the $Z\rightarrow\tau\tau$ decay, therefore the reconstructed mass of the $\tau$ pair, $m_{\tau\tau}$ is required to be below 50~\GeV; $m_{\tau\tau}$ is computed using the collinear approximation method~\cite{Plehn:1999xi}.
In addition, the azimuthal separation between the Higgs-candidate lepton pair and the \pmet, $\Delta\phi_{\ell_0\ell_1,\,\pmet}$, is required to be above 0.4~rad.
The final selection of these variables is optimised in order to maximise the signal significance. Figure~\ref{fig:DPhillMET_1SFOS} illustrates the discriminating power of the $\Delta\phi_{\ell_0\ell_1,\,\pmet}$ variable for 1-SFOS events after the preselection. This variable reduces the ZZ* contribution which becomes the dominant background source after the $b$-tagged jet veto and $m_{\ell_0 \ell_1}$ selection are applied.
Figure~\ref{fig:MET_2SFOS} shows the  $E_{\textrm{T}}^{\textrm{miss}}$ distribution for 2-SFOS events after the preselection.
In order to be orthogonal to the $H\rightarrow ZZ^{\ast}$ analysis of Ref.~\cite{HIGG-2016-25}, 2-SFOS events are required to have an invariant mass of the four-lepton system, $m_{4\ell}$, above 140~\GeV. The full event selection for \ZH is summarised 
in Table~\ref{tab:SRsSelection}.

\begin{table}
\begin{center}
\caption{Summary of the $Z$-depleted fit regions.}
\label{tab:ZdeplBDTbins}
\resizebox{\textwidth}{!}{
\begin{tabular}{c|c|c|c|c|c}
\hline  \hline
Bin 1 & Bin 2 & Bin 3 & Bin 4 & Bin 5 & Bin 6 \\
\hline
\multicolumn{2}{c|}{$0.2\leq \mathrm{BDT}_{t\bar{t}} <0.3$} & \multicolumn{2}{c|}{$0.3\leq \mathrm{BDT}_{t\bar{t}} <0.45$} & \multicolumn{2}{c}{$0.45\leq \mathrm{BDT}_{t\bar{t}}$}\\
$0.15\leq \mathrm{BDT}_{WZ}<0.35$ & $0.35\leq \mathrm{BDT}_{WZ}$ &$0.15\leq \mathrm{BDT}_{WZ}<0.35$ & $0.35\leq \mathrm{BDT}_{WZ}$ & $0.15\leq \mathrm{BDT}_{WZ}<0.35$ & $0.35\leq \mathrm{BDT}_{WZ}$\\
\hline \hline
\end{tabular}
}
\end{center}
\end{table}

The total efficiency times acceptance of this selection for the process \WH with subsequent \HWW decay is about $0.073\%$ while for \ZH it is about $0.026\%$. These numbers are given with respect to all $W$ and $Z$ decays.
\section{Background estimation}
\label{sec:background}
 
The main background contamination originates from processes with the same final state as the signal, namely diboson production ($WZ/W\gamma^{\ast}$, $ZZ^{\ast}$), top-quark processes with three or four prompt leptons such as $t\bar{t}V$,
and triboson production. Other relevant background contributions arise from processes, such as $t\bar{t}$ or $Z+$jets, where the reconstructed leptons either originate from  non-prompt  decays of heavy-flavour hadrons or from jets misidentified as leptons.

Two dedicated regions, hereafter named control regions  (CRs), are used to estimate the normalisation factors (NFs) of the main prompt background processes by fitting the expected yields from simulation to data: $WZ/W\gamma^{\ast}$ for the \WH channel and $ZZ^{\ast}$ for the \ZH channel in the 2-SFOS SR. In the 1-SFOS SR, $ZZ^{\ast}$ is estimated purely from simulation.
 
\begin{figure}
\centering
\subfloat[]{
\includegraphics[width=0.5\linewidth]{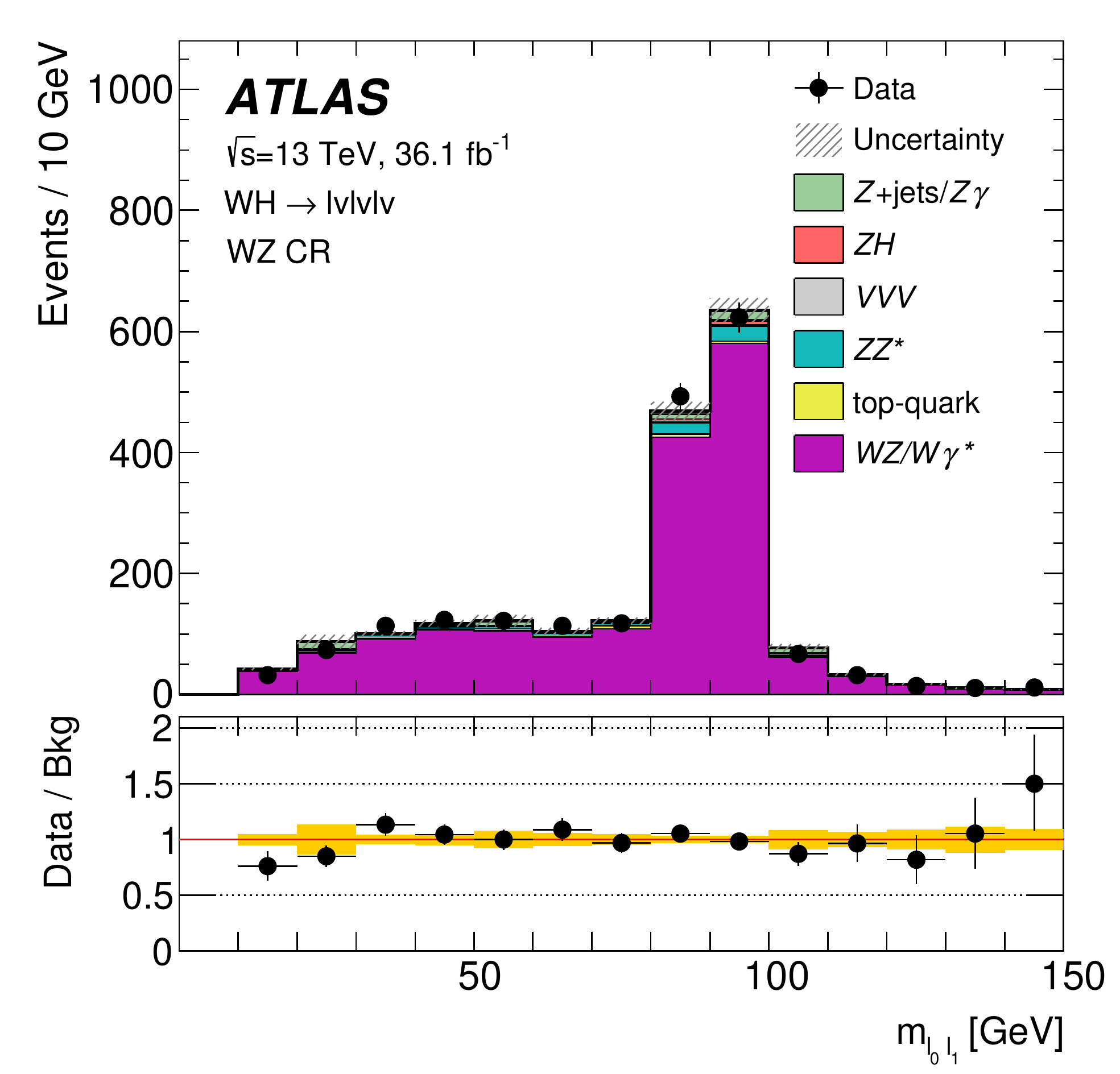}\label{fig:WZCR}
}
\subfloat[]{
\includegraphics[width=0.5\linewidth]{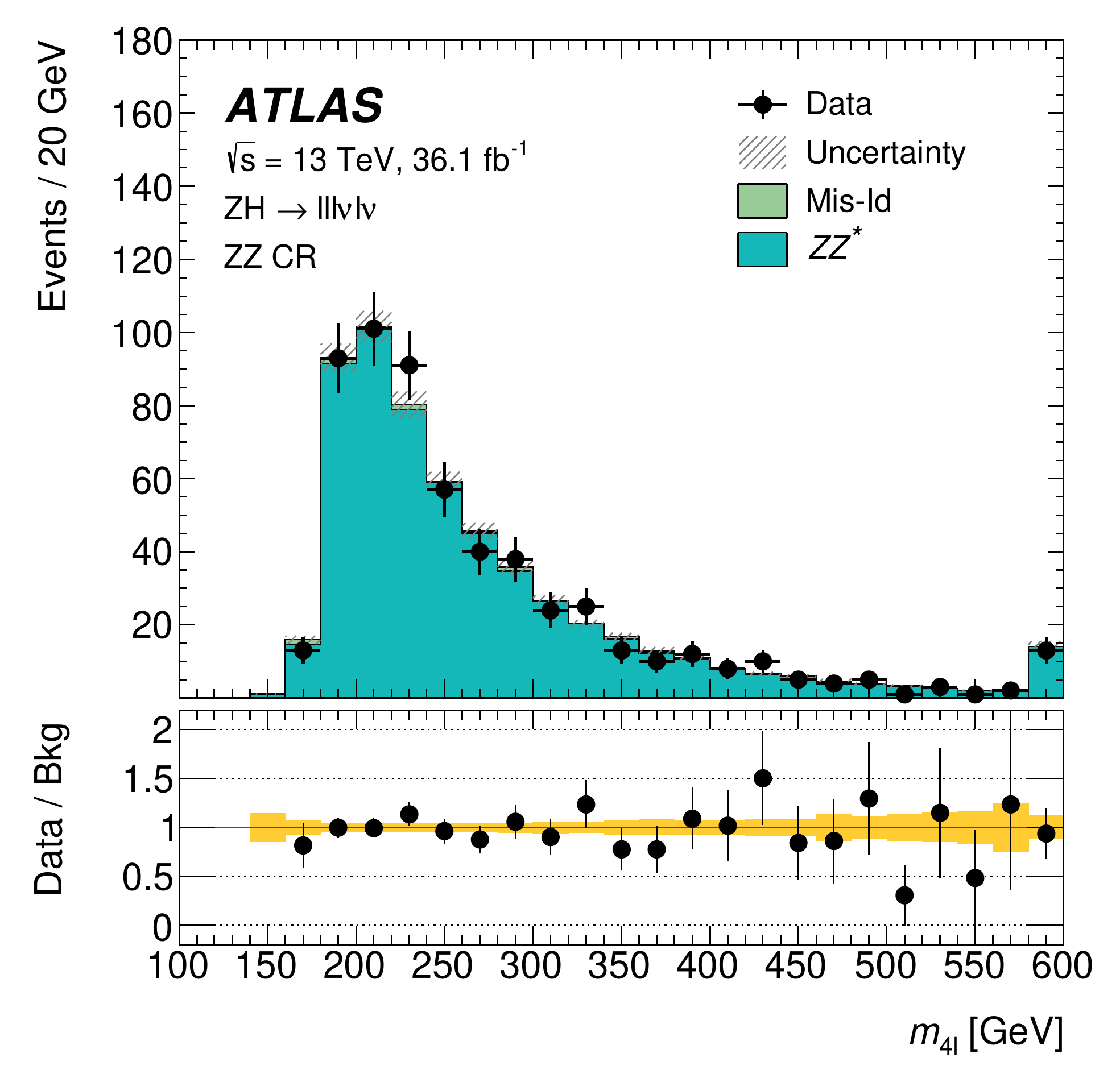} \label{fig:ZZCR}
}\newline
\subfloat[]{
\includegraphics[width=0.5\linewidth]{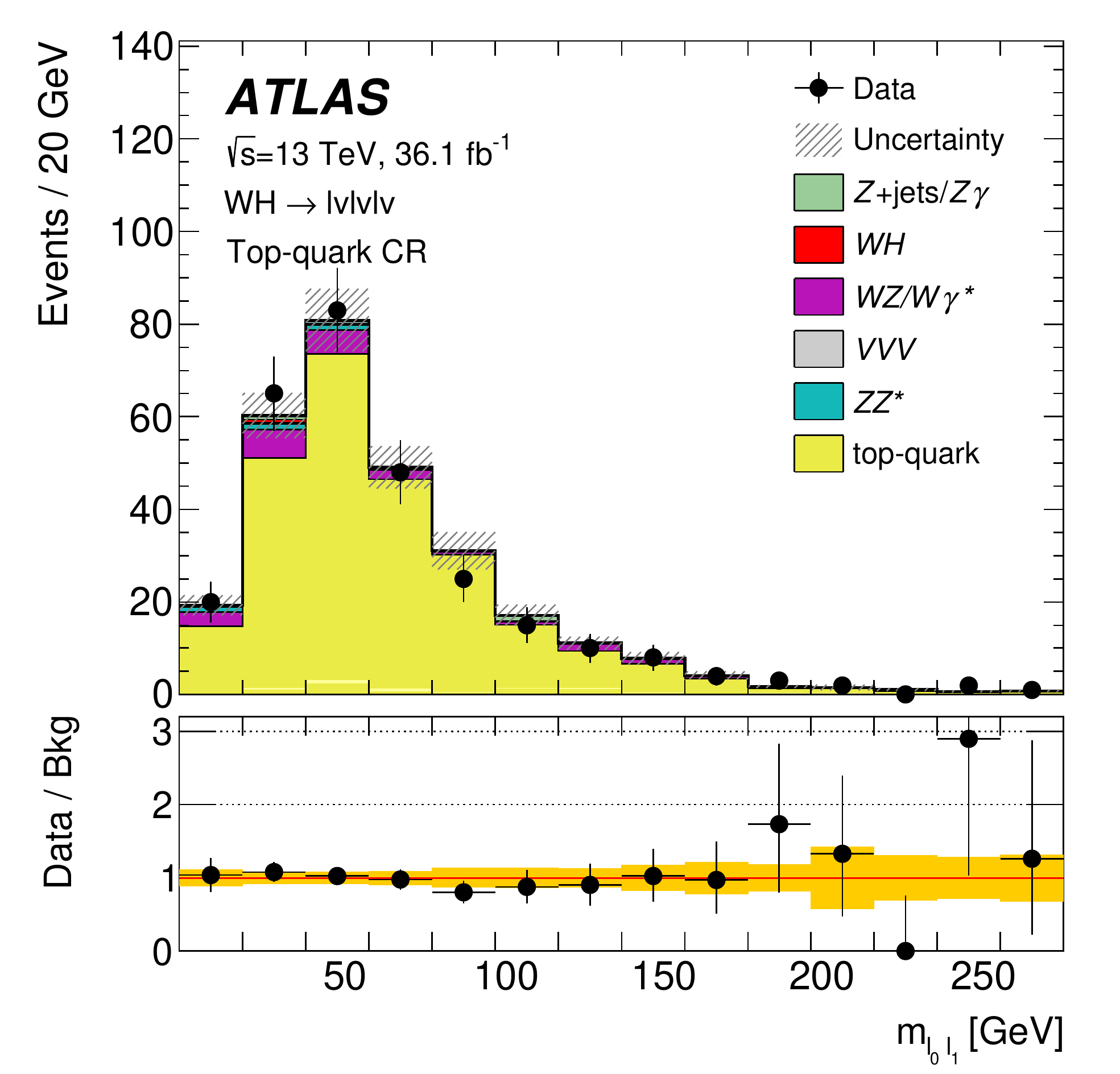}\label{fig:topCR}
}
\caption{Post-fit distributions of the dilepton invariant mass $m_{\ell_0\ell_1}$ in the $WZ$ CR~\protect\subref{fig:WZCR}  and  of the four-lepton invariant mass $m_{4\ell}$
in the $ZZ$ CR~\protect\subref{fig:ZZCR}. For the latter the last bin contains overflow events. Post-fit distributions of the dilepton invariant mass $m_{\ell_0\ell_1}$ is shown also
in the top-quark CR~\protect\subref{fig:topCR}.
The hatched band in the upper panel and the shaded band in the lower panel show the total statistical and experimental systematic uncertainty in background predictions. The Mis-Id background is estimated with the data-driven technique described in Section~\ref{sec:background}.
}
\end{figure}

The CRs are made orthogonal to the corresponding SRs  by inverting various selection criteria with respect to the SR definitions.
The $WZ$ CR is defined by reversing the $Z$-veto in the $Z$-dominated \WH signal region.
To improve the purity of the $WZ$ CR, the minimum $\met$ is increased from 30~\gev\ to 50~\gev.
The $ZZ$ CR is defined by inverting the $m_{\ell_0\ell_1}$ requirement defined in the \ZH 2-SFOS SR. In order to increase the
number of events, the $\met$, $\Delta \phi_{\ell_0\ell_1}^{\mathrm{boost}}$ , $\pt^{4\ell}$ , and $m_{4\ell}$ requirements are removed.
Table~\ref{tab:CRsSelection} summarises the event selection for the $WZ$ and $ZZ$ CRs and the NFs for the background processes, obtained from the fit described in Section~\ref{sec:results}. The NFs are hereby completely dominated by the CR statistics and their values do not change when only the control regions are used in the fit.
\begin{table}
\begin{center}
\caption{ Definition of control regions in the \WH and \ZH analyses. Selections indicated in boldface font are designed to ensure the control region (CR) is orthogonal to the relevant SR. In the last line  normalisation factors  which scale the corresponding yields in the signal region are shown with their  uncertainties, including both the statistic and the systematic uncertainties.
}
\begin{tabular}{l||cc|c}
\hline \hline
Channel (Category) & \multicolumn{2}{c|}{ $WH$ ($Z$-dominated and $Z$-depleted) } &  \multicolumn{1}{c}{$ZH$ (2-SFOS)} \\
\hline
CR                                       & $WZ$                          & Top-quark             & $ZZ$                    \\
\hline
Number of leptons                                    &3                           &3          &4                       \\
Total lepton charge                                  &$\pm1$                      &$\pm1$       &0                    \\
Number of SFOS                                       &2 or 1                      & $\leq2$      &2                    \\
Number of jets                                    &$\le1$                     &$\ge1$           &$\le1$               \\
Number of $b$-jets                                   &0                        &{$\mathbf{\ge1}$}     &0            \\
\met\ [\gev]                                        &$>50$                       &$>50$ (2 or 1 SFOS)     &---                \\
$|m_{\ell\ell} - m_{Z}|$ [\gev]              &$\mathbf{<25}$         &$>25$ (2 or 1 SFOS)      &$<10 \ (m_{\ell_2\ell_3})$         \\
$m_{\ell_0\ell_1}$ [\gev]                       & ---                   &  ---             &$\mathbf{>55}$                \\
$m_{\ell\ell}$ (min. SFOS) [\gev]                            & $>12$                   &$>12$ (2 or 1 SFOS)   & ---              \\
\hline \hline
Normalisation factors                   & 0.99$\pm$0.05              &  0.97$\pm$0.08        &  1.13$\pm$0.06       \\
\hline\hline
\end{tabular}
\label{tab:CRsSelection}
\end{center}
\end{table}
Figures~\ref{fig:WZCR}  and~\ref{fig:ZZCR} show the distributions of  $m_{\ell_0\ell_1}$ in the $WZ$ CR and the invariant mass of the four leptons in the $ZZ$ CR, as obtained from the final fit in the statistical analysis (post-fit) described in Section~\ref{sec:results}.
 
The background contributions  with misidentified leptons are estimated using different techniques. The top-quark background in the $WH$ channels is normalised  using a CR (top-quark CR) defined by requiring at least one $b$-tagged jet.
To improve the purity of the top-quark CR, the minimum $\met$ is increased from 30~\gev\ to 50~\gev\ if at least one SFOS pair is present in the event. Processes with one misidentified lepton ($t\bar{t}$ and $Wt$) constitute  94$\%$ of the top-quark CR, the remaining events contain three prompt leptons from $t\bar{t}W$ decays. The full selection requirements applied to define the top-quark CR and the measured  NF are also summarised in Table~\ref{tab:CRsSelection}. Figure~\ref{fig:topCR}  shows the $m_{\ell_0\ell_1}$ distribution in the top-quark CR  as obtained from the final fit.
 
A data-driven technique is used to estimate the $Z$+jets and $Z\gamma$ contribution in both the \WH and \ZH channels and the contributions from $WZ/W\gamma^{\ast}$ and top-quark processes in the \ZH channel. These contributions typically have one misidentified lepton ($Z$+jets and $Z\gamma$ in the \WH channel, and $WZ/W\gamma^{\ast}$ and $tZ/t\bar{t}W$ in the \ZH channel) or two misidentified leptons ($Z$+jets and $tW/t\bar{t}$ in the \ZH channel).
A control sample where one or two of the lepton	candidates fail to meet the nominal identification or isolation criteria but satisfy looser identification criteria, referred to as anti-identified leptons, is used. The contribution from misidentified leptons in the SR is then obtained by scaling the number of events in the control sample by  extrapolation factors  measured in a data sample enriched in $Z$+jets events.  The latter is obtained by selecting events with two prompt leptons from a $Z$ boson decay and a loosely identified lepton considered to be the misidentified lepton candidate. The extrapolation factors are defined as the ratio of the number of misidentified lepton candidates that pass the nominal identification criteria to the number that pass the anti-identification criteria.
In both the control sample and the data samples enriched in $Z$+jets events, the contribution from background events not estimated with this method is subtracted using MC expectations. Details of this method can be found in Ref.~\cite{ATLAS:2014aga}.  The  uncertainty in the data-driven background processes described in this section includes  the statistical uncertainty in the $Z$+jets enriched sample, the uncertainty from $Z$+jets MC modelling, and the theory uncertainty from the subtraction of other processes. For the \ZH channel, which can have events with two prompt leptons and two misidentified leptons, the uncertainty  in the extrapolation from the  control sample to the SRs is also included.

\section{Systematic uncertainties}
\label{sec:systematics}
The systematic uncertainties can be categorised into those arising
from experimental sources and those from theoretical sources. The dominant
experimental uncertainties come from the misidentification of leptons
(see Section~\ref{sec:background}),
the mismodelling of the impact-parameter significance, the \textit{b}-tagging
efficiency~\cite{PERF-2016-05} and the jet energy scale and
resolution~\cite{PERF-2016-04}. Other sources of uncertainty
are due to the modelling of pile-up, the calibration of the missing transverse momentum
measurement~\cite{PERF-2016-07}, and the luminosity
measurement.
The uncertainty in the combined 2015+2016 integrated luminosity is 2.1\%. It is derived from the calibration of the luminosity scale using $x$-$y$ beam-separation scans, following a methodology similar to that detailed in Ref.~\cite{DAPR-2013-01}, and using the LUCID-2 detector for the baseline luminosity measurements \cite{LUCID2}.
 
The impact of the uncertainties on lepton energy
(momentum) scale and resolution, and identification and
isolation
criteria~\cite{PERF-2017-03,PERF-2017-01,PERF-2015-10}
is negligible. The experimental uncertainties are
varied in a correlated way across all background processes and all
signal- and control-region bins,  so that uncertainties in the
extrapolation from control to signal regions are correctly
propagated.
The luminosity uncertainty is only applied to background processes
that are normalised to theoretical predictions, and to the Higgs boson
signal.
The theoretical uncertainties are evaluated by comparing nominal and  alternative event generators and
UEPS models as described in Section~\ref{sec:DataMC} and by varying PDF sets and the QCD renormalisation and
factorisation scales. The uncertainty due to the PDF choice for the
signal process is $1\%$ while the uncertainty obtained by using \textsc{Herwig} as an alternative parton shower model is $3 - 10\%$, depending on the signal region.
All uncertainties are propagated through the full analysis chain and treated
as being bin-dependent and region-dependent, i.e.\ potentially modifying not only the normalisation but also the shape of the BDT output distributions. 
Whenever the influence on the shape is found to be negligible, as in the case of the PDF and scale variations, only the normalisation uncertainties are used.
A list of the systematic uncertainty sources and their impact on the cross-section measurement
are shown later in Section~\ref{sec:results}.

\section{Results}
\label{sec:results}
A binned likelihood function is constructed as a product of Poisson
probability terms over the eleven bins of the different SRs defined in
Section~\ref{sec:analysis}. The function has two independent scaling
parameters: the signal strength parameter $\mu$, defined
as the ratio of the measured signal yield to that predicted by the SM,
for each of the \WH and the \ZH processes. Additionally, one Poisson
probability term is added for each CR to determine simultaneously the normalisation of the corresponding background processes.
Systematic uncertainties enter as nuisance parameters in the likelihood
function and their correlations are taken into account. The final
results are obtained using the profile likelihood method~\cite{Cowan:2010js}.
The resulting post-fit prediction and data yields in the four SRs are shown in Table~\ref{tab:postfit_yields}.
 

\begin{table}
\centering
\caption{\label{tab:postfit_yields}Post-fit predictions and data yields in the
four SRs. The uncertainties include those from the sample
statistics, and the theoretical and experimental systematic
uncertainties. The sum of the single contributions may differ from
the total value due to rounding. Moreover, the total uncertainty
differs from the sum in quadrature of the single-process
uncertainties due to the correlations.  ``Other Higgs'' contains
Higgs production mechanisms and decay processes different from $VH$
and $H\to WW^{\ast}$ respectively, except for the $H \to ZZ^{\ast}$
contribution which is included in the ``$ZZ^{\ast}$'' row. The corresponding pre-fit predictions for the background processes differ only by the normalization factors listed in Table~\ref{tab:CRsSelection}.}
\begin{tabular}{lS[table-format=2.2,table-number-alignment=right]@{$\,\pm\,$}
S[table-format=1.2,table-number-alignment=left]
S[table-format=2.2,table-number-alignment=right]@{$\,\pm\,$}
S[table-format=1.1,table-number-alignment=left]
S[table-format=1.2,table-number-alignment=right]@{$\,\pm\,$}
S[table-format=1.2,table-number-alignment=left]
S[table-format=1.2,table-number-alignment=right]@{$\,\pm\,$}
S[table-format=1.2,table-number-alignment=left]}
\dbline
Process & \multicolumn{4}{c}{$WH$} & \multicolumn{4}{c}{$ZH$}  \tabularnewline
& \multicolumn{2}{c}{$Z$-dominated}  &  \multicolumn{2}{c}{$Z$-depleted}         &  \multicolumn{2}{c}{1-SFOS} &\multicolumn{2}{c}{2-SFOS}\tabularnewline
\sgline
$WH$  &    11 & 6             &  5.8 & 2.8                 &\multicolumn{4}{c}{---}     \tabularnewline
$ZH$  &    1.1 & 0.6               &   0.61 & 0.34                  &  3.3 & 1.7 &          1.8   &  0.9  \tabularnewline
\sgline
$WZ/W\gamma^{\ast}$                        & 40.1 & 2.8                & 1.7 & 0.5  &\multicolumn{4}{c}{---}     \tabularnewline
$ZZ^{\ast}$                         & 2.4 & 1.1                  & 0.27 & 0.09  & 0.14 & 0.14 & 1.2 & 0.3 \tabularnewline
$VVV$                         & 1.5 & 0.1                  & 0.71 & 0.11  & 0.32 & 0.05 & 0.20 & 0.03 \tabularnewline
$tV/t\bar{t}V$                 & 0.14 & 0.03 &            0.13 & 0.03 & 0.04 & 0.02& 0.03 & 0.01 \tabularnewline
Other top-quark                & 8.4 & 2.6  & 1.9 & 0.8 & \multicolumn{4}{c}{---}     \tabularnewline
Other Higgs                    & 0.31 & 0.03 & 0.06 & 0.01 &
\multicolumn{2}{c}{<0.01}
& 0.04 & 0.01\tabularnewline
Misid. leptons              &   9.7   & 3.4              &   \multicolumn{2}{c}{<0.1}      &  0.19  & 0.08 &              0.36   &  0.12   \tabularnewline
\sgline
Total  background        &   62 & 5              & 4.7 &  1.0                                       &  0.65 & 0.17 &                1.8  &  0.3  \tabularnewline
Observed  &  \multicolumn{2}{l}{76}  &  \multicolumn{2}{l}{10} &  \multicolumn{2}{l}{5} & \multicolumn{2}{l}{2}  \tabularnewline
\dbline
\end{tabular}
\end{table}
 
\begin{figure}
\centering
\subfloat[]{
\includegraphics[width=0.5\linewidth]{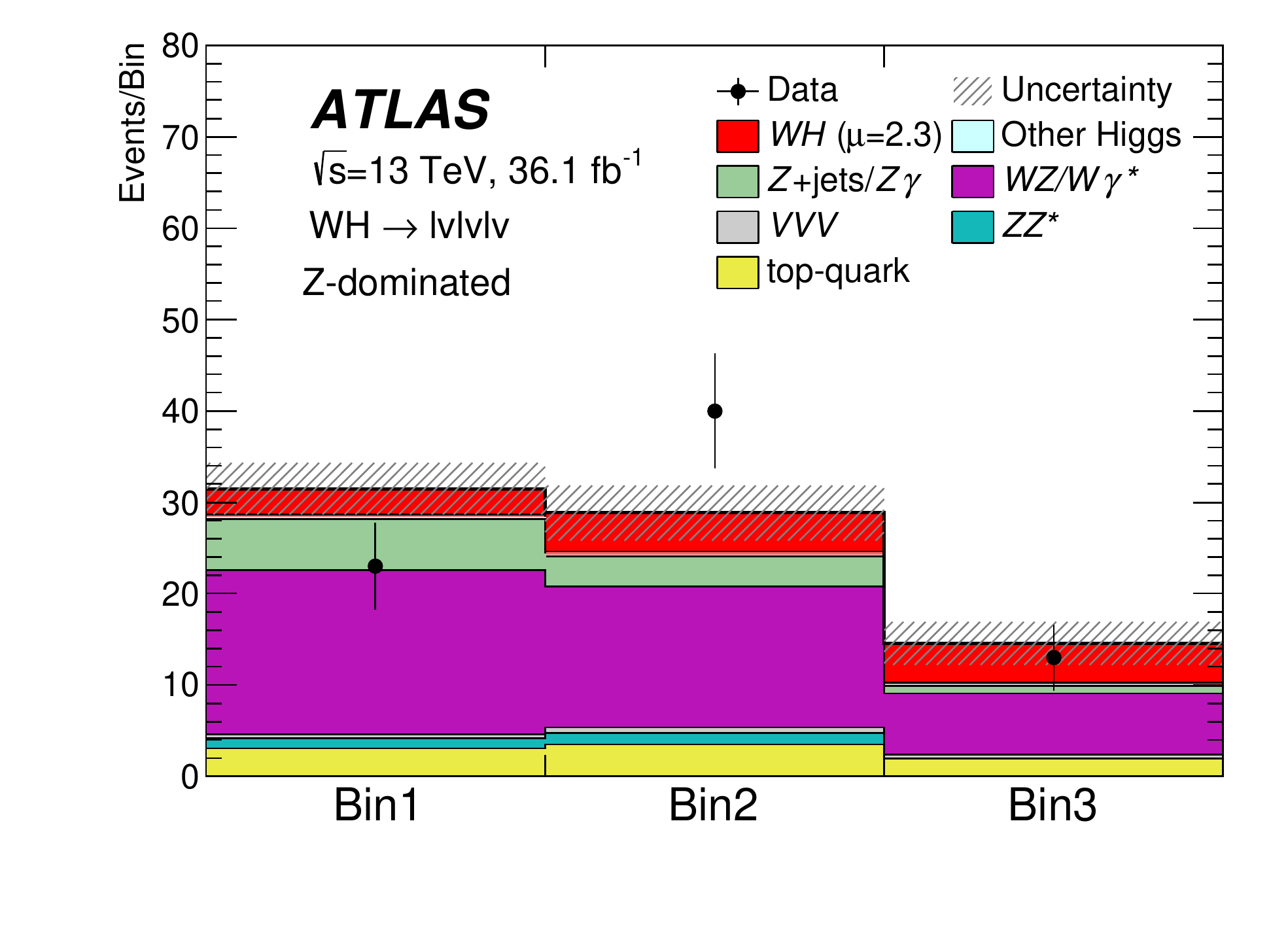}\label{fig:SR12_3bins}
}
\subfloat[]{
\includegraphics[width=0.5\linewidth]{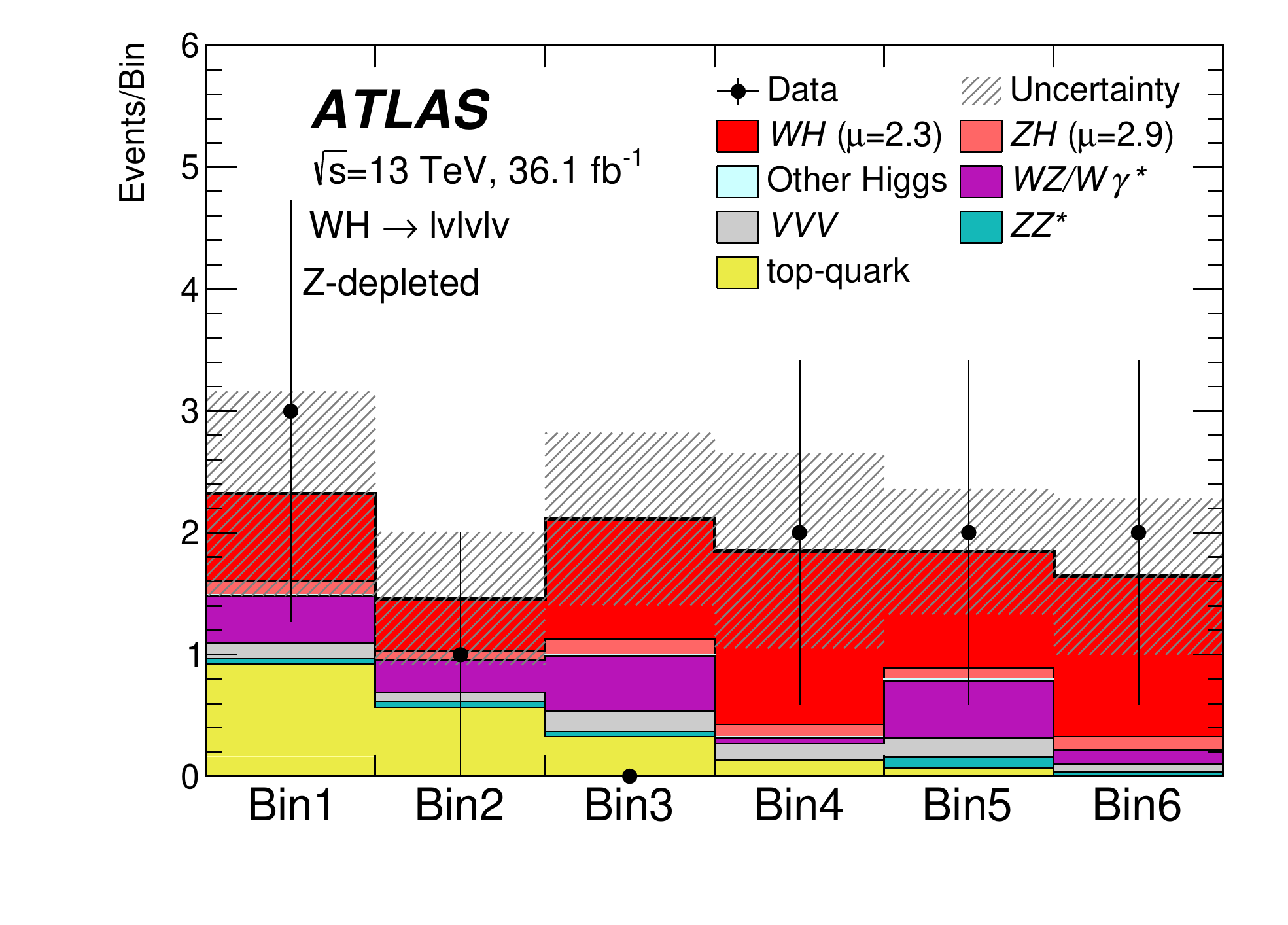}\label{fig:SR3_6bins}
}
\caption{\label{fig:SR_WH}Post-fit BDT-score distributions \protect\subref{fig:SR12_3bins} in the $WH$ $Z$-dominated SR
and \protect\subref{fig:SR3_6bins} for the two-dimensional grid in \bdttop and \bdtwz in the $WH$ $Z$-depleted SR.
The shaded band includes statistical and systematic uncertainties on both signal and background as estimated by the fit.}
\end{figure}

\begin{figure}
\centering
\includegraphics[width=0.5\linewidth]{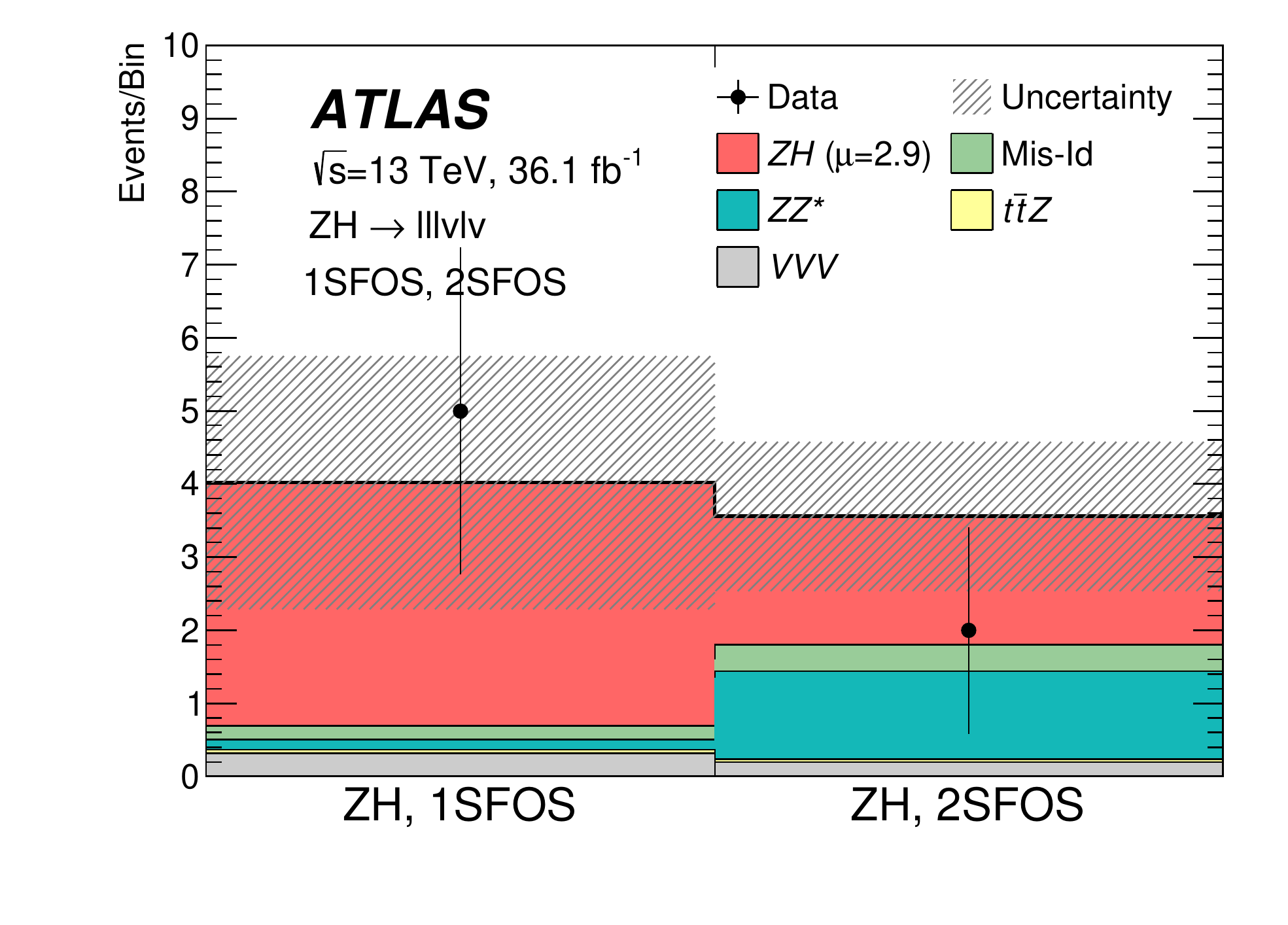}
\caption{\label{fig:ZH_EvtYieldFinal} Post-fit event yields in the
$ZH$ 1-SFOS and 2-SFOS SRs. The shaded band includes statistical
and systematic uncertainties on both signal and background as estimated by the fit.}
\end{figure}

Figure~\ref{fig:SR_WH} shows the post-fit distribution of BDT scores in the $WH$ $Z$-dominated and $Z$-depleted SRs.
The post-fit event yields in the $ZH$ 1-SFOS and 2-SFOS SRs are shown in Figure~\ref{fig:ZH_EvtYieldFinal}.
 
 
The measured signal strengths for the $WH$ and $ZH$ production modes in the $\hww$ decay channel are simultaneously determined to be
\begin{eqnarray*}
\mu_{WH} &=&  2.3^{+1.1}_{-0.9} (\mathrm{stat.})_{-0.36}^{+0.49}  (\mathrm{exp~ syst.}) _{-0.33}^{+0.41} (\mathrm{theo~ syst.})  = 2.3^{+1.2}_{-1.0}, \\
\mu_{ZH} &=&  2.9^{+1.7}_{-1.3} (\mathrm{stat.}) _{-0.28}^{+0.54}(\mathrm{exp~ syst.}) _{-0.27}^{+0.66}(\mathrm{theo~ syst.}) = 2.9^{+1.9}_{-1.3},
\end{eqnarray*}
compared to the expected results of
$\mu_{WH}=1.0^{+0.86}_{-0.69}(\mathrm{stat.})^{+0.40}_{-0.30}(\mathrm{exp~syst.})^{+0.32}_{-0.26}(\mathrm{theo~ syst.}) =1.0^{+1.0}_{-0.8}$ and
$\mu_{ZH}=1.0^{+1.3}_{-0.9}(\mathrm{stat.}) ^{+0.41}_{-0.20}(\mathrm{exp~syst.}) ^{+0.32}_{-0.16}(\mathrm{theo~ syst.}) =1.0^{+1.4}_{-0.9}$.
The observed (expected) significances of $WH$ and $ZH$ production modes
are 2.6 (1.3) standard deviations and 2.8 (1.2) standard
deviations above the SM background, including other Higgs-boson processes, respectively. When determining the
significance for $WH$ production, the $ZH$ signal-strength parameter is left floating in the fit, and vice versa.
The combination of the $WH$ and $ZH$ channels leads to an observed (expected)
significance for the combined $VH$
production mode of  4.1 (1.9) standard deviations above the SM
background, including other Higgs-boson processes. The $p$-value with respect to the value predicted by the SM corresponds to about two standard deviations. The validity of the asymptotic approximation used in deriving these results was tested using pseudo-experiments.
The combined signal strength for $VH$ is measured to be
\begin{eqnarray*}
\mu_{VH} &=&  2.5^{+0.8}_{-0.7} (\mathrm{stat.}) _{-0.23}^{+0.30}(\mathrm{exp~ syst.}) _{-0.26}^{+0.37}  (\mathrm{theo~ syst.})  = 2.5^{+0.9}_{-0.8} \\
\end{eqnarray*}
with an expected signal strength of $\mu_{VH}=1.0^{+0.66}_{-0.54}(\mathrm{stat.})^{+0.26}_{-0.17}(\mathrm{exp~syst.})^{+0.22}_{-0.17}(\mathrm{theo~ syst.}) =1.0^{+0.7}_{-0.6}$.

The cross-section times branching-fraction values, $\sigma_{WH} \cdot \mathcal{B}_{\hww}$ and $\sigma_{ZH} \cdot \mathcal{B}_{\hww}$, are simultaneously determined to be:
 
\begin{eqnarray*}
\sigma_{WH} \cdot \mathcal{B}_{H \to WW^{\ast}} &=& 0.67^{+0.31}_{-0.27} (\mathrm{stat.})^{+0.14}_{-0.11}(\mathrm{exp~ syst.}) ^{+0.11}_{-0.09}(\mathrm{theo~ syst.})~\mathrm{pb},\\
\sigma_{ZH} \cdot \mathcal{B}_{H \to WW^{\ast}} &=& 0.54^{+0.31}_{-0.24}(\mathrm{stat.})^{+0.10}_{-0.05}(\mathrm{exp~ syst.}) ^{+0.11}_{-0.05}(\mathrm{theo~ syst.})~\mathrm{pb}.
\end{eqnarray*}
 
\begin{figure}
\centering
\includegraphics[width=0.4\linewidth]{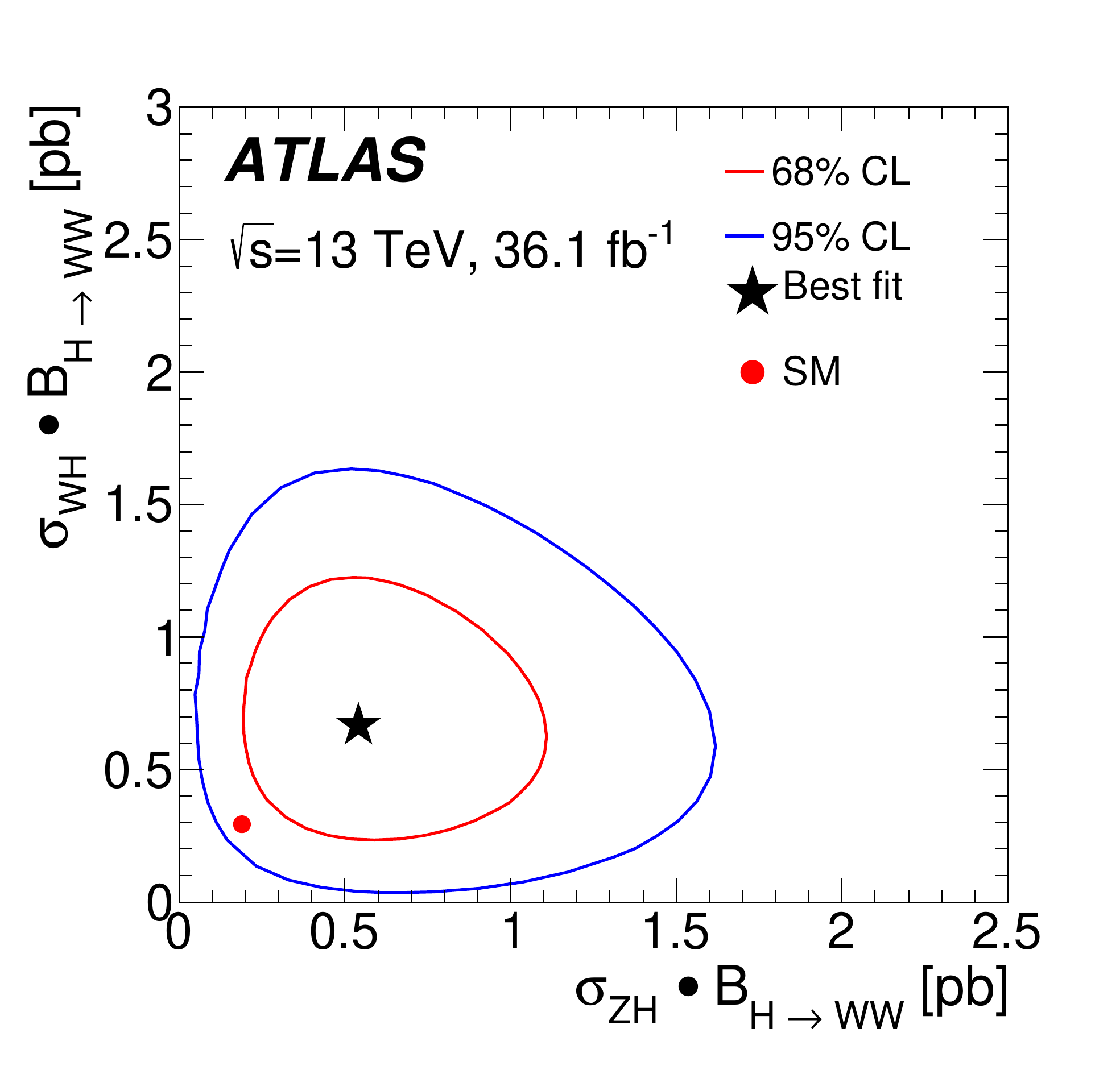}
\caption{\label{fig:2dsigma}Two-dimensional likelihood contours of
$\sigma_{WH} \cdot \mathcal{B}_{\hww}$ vs. $\sigma_{ZH} \cdot \mathcal{B}_{\hww}$ for $68\%$ and
$95\%$ confidence level (CL) compared with the prediction from the Standard Model.}
\end{figure}

The main contributions to the uncertainties in $\sigma_{WH} \cdot \mathcal{B}_{\hww}$ and $\sigma_{ZH} \cdot \mathcal{B}_{\hww}$ are summarised in Table~\ref{tab:sys_breakdown}.
The predicted cross-section times branching-fraction values are $ 0.293 \pm 0.007$~pb and $0.189\pm 0.007$~pb for $WH$ and $ZH$~\cite{deFlorian:2016spz}, respectively.
The 68\% and 95\% confidence level two-dimensional contours of
$\sigma_{WH} \cdot \mathcal{B}_{H \to WW^{\ast}}$ and $\sigma_{ZH}
\cdot \mathcal{B}_{H \to WW^{\ast}}$ are shown in
Figure~\ref{fig:2dsigma}. 
The corresponding one-dimensional results, where the other parameter is left floating in the fit,
are consistent with the SM predictions within $1.3\sigma$ for $WH$ and $1.5\sigma$ for $ZH$.
 
\begin{table}
\centering
\caption{\label{tab:sys_breakdown}Breakdown of the main contributions to the total uncertainty in $\sigma_{WH} \cdot \mathcal{B}_{\hww}$ (left) and $\sigma_{ZH}  \cdot \mathcal{B}_{\hww}$ (right). The individual sources of systematic uncertainties are grouped together. The sum in quadrature of the individual components differs from the total uncertainty due to correlations between the components. Systematic uncertainties that affect the shape of the fitted distribution are indicated by an asterix.}
 
\begin{minipage}[b]{80mm}
\scalebox{0.8}{
\begin{tabular}{lrr}
\hline\hline
Source          & $\Delta\sigma_{WH} \cdot \mathcal{B}_{\hww}$ $[\%]$ \\
\hline\hline
Data statistics  in SR                        & $43$ \\ 
Data statistics  in CR                        & $6$ \\ 
\hline
Theoretical uncertainties                       & $16$ \\ 
\hspace*{4mm}$WZ/W\gamma^{\ast}$(*)             & $12$ \\ 
\hspace*{4mm}Top-quark(*)                       & $8$ \\ 
\hspace*{4mm}$WH$ signal(*)                     & $4$ \\ 
\hspace*{4mm}$ZZ^{\ast}$                         & $2$ \\ 
Experimental uncertainties                      & $12$ \\ 
\hspace*{4mm}Impact parameter mismodelling      & $8$ \\ 
\hspace*{4mm}Misidentified leptons              & $8$ \\ 
\hspace*{4mm}$b$-tagging                        & $1$ \\ 
MC statistics                                   & $9$ \\ 
Luminosity                                      & $3$ \\ 
\hline
TOTAL                                           & $49$ \\ 
\hline\hline
\end{tabular}
}
\end{minipage}
\begin{minipage}[b]{80mm}
\scalebox{0.8}{
\begin{tabular}{lrr}
\hline\hline
Source          & $\Delta\sigma_{ZH} \cdot \mathcal{B}_{\hww}$ $[\%]$  \\
\hline\hline
Data statistics  in SR                         & $50$ \\ 
Data statistics  in CR                               & $<1$\\ 
\hline
Theoretical uncertainties                 & $15$\\ 
\hspace*{4mm}$ZH$ signal             & $14$ \\ 
\hspace*{4mm}Top-quark                    & $1$ \\ 
\hspace*{4mm}$WZ/W\gamma^{\ast}$                         & $<1$ \\ 
\hspace*{4mm}$ZZ^{\ast}$                         & $<1$\\ 
Experimental uncertainties                    & $7$ \\ 
\hspace*{4mm}Misidentified leptons            & $3$ \\ 
\hspace*{4mm}$b$-tagging                      & $1$ \\ 
\phantom{Impact parameter mismodelling} & \phantom{0}\\
MC statistics                                 & $11$\\ 
Luminosity                                    & $2$ \\ 
\hline
TOTAL                                         & $54$ \\ 
\hline\hline
\end{tabular}
}
\end{minipage}
\end{table}

\FloatBarrier
\section{Conclusion}
Measurements of the inclusive production of a Higgs boson in association with
a $W$ or $Z$ boson are presented.
Results of the $WH$ and $ZH$ production cross section times branching fraction for $H \rightarrow WW^{\ast}$ are obtained selecting the leptonic decays
$W \rightarrow \ell\nu$ and $Z \rightarrow \ell\ell$, with $\ell = e,\ \mu$,
in $\sqrt{s} = 13\,\mathrm{TeV}$ proton--proton collisions
recorded with the ATLAS detector at the LHC
in a data sample corresponding to an integrated luminosity of $36.1\,\mathrm{fb}^{-1}$.
The products of the $\hww$ branching fraction times the $WH$ and $ZH$ cross sections are measured to be
$0.67^{+0.31}_{-0.27}$(stat.)$^{+0.18}_{-0.14}$(syst.)~pb and $0.54^{+0.31}_{-0.24}$(stat.)$^{+0.15}_{-0.07}$(syst.)~pb, respectively,
compatible with the Standard Model predictions
of $ 0.293 \pm 0.007$~pb for $WH$ and $0.189\pm 0.007$~pb for $ZH$
within up to $1.5\sigma$. The $WH$ channel with $H \rightarrow
WW^{\ast}$ is purely sensitive to the $HW$ coupling and this
measurement is to date the most precise measurement of this channel.
 

\section*{Acknowledgements}
 
 
We thank CERN for the very successful operation of the LHC, as well as the
support staff from our institutions without whom ATLAS could not be
operated efficiently.
 
We acknowledge the support of ANPCyT, Argentina; YerPhI, Armenia; ARC, Australia; BMWFW and FWF, Austria; ANAS, Azerbaijan; SSTC, Belarus; CNPq and FAPESP, Brazil; NSERC, NRC and CFI, Canada; CERN; CONICYT, Chile; CAS, MOST and NSFC, China; COLCIENCIAS, Colombia; MSMT CR, MPO CR and VSC CR, Czech Republic; DNRF and DNSRC, Denmark; IN2P3-CNRS, CEA-DRF/IRFU, France; SRNSFG, Georgia; BMBF, HGF, and MPG, Germany; GSRT, Greece; RGC, Hong Kong SAR, China; ISF and Benoziyo Center, Israel; INFN, Italy; MEXT and JSPS, Japan; CNRST, Morocco; NWO, Netherlands; RCN, Norway; MNiSW and NCN, Poland; FCT, Portugal; MNE/IFA, Romania; MES of Russia and NRC KI, Russian Federation; JINR; MESTD, Serbia; MSSR, Slovakia; ARRS and MIZ\v{S}, Slovenia; DST/NRF, South Africa; MINECO, Spain; SRC and Wallenberg Foundation, Sweden; SERI, SNSF and Cantons of Bern and Geneva, Switzerland; MOST, Taiwan; TAEK, Turkey; STFC, United Kingdom; DOE and NSF, United States of America. In addition, individual groups and members have received support from BCKDF, CANARIE, CRC and Compute Canada, Canada; COST, ERC, ERDF, Horizon 2020, and Marie Sk{\l}odowska-Curie Actions, European Union; Investissements d' Avenir Labex and Idex, ANR, France; DFG and AvH Foundation, Germany; Herakleitos, Thales and Aristeia programmes co-financed by EU-ESF and the Greek NSRF, Greece; BSF-NSF and GIF, Israel; CERCA Programme Generalitat de Catalunya, Spain; The Royal Society and Leverhulme Trust, United Kingdom.
 
The crucial computing support from all WLCG partners is acknowledged gratefully, in particular from CERN, the ATLAS Tier-1 facilities at TRIUMF (Canada), NDGF (Denmark, Norway, Sweden), CC-IN2P3 (France), KIT/GridKA (Germany), INFN-CNAF (Italy), NL-T1 (Netherlands), PIC (Spain), ASGC (Taiwan), RAL (UK) and BNL (USA), the Tier-2 facilities worldwide and large non-WLCG resource providers. Major contributors of computing resources are listed in Ref.~\cite{ATL-GEN-PUB-2016-002}.
 

\clearpage

\printbibliography
 
\clearpage 
 
\begin{flushleft}
{\Large The ATLAS Collaboration}

\bigskip

G.~Aad$^\textrm{\scriptsize 101}$,    
B.~Abbott$^\textrm{\scriptsize 128}$,    
D.C.~Abbott$^\textrm{\scriptsize 102}$,    
O.~Abdinov$^\textrm{\scriptsize 13,*}$,    
A.~Abed~Abud$^\textrm{\scriptsize 70a,70b}$,    
K.~Abeling$^\textrm{\scriptsize 53}$,    
D.K.~Abhayasinghe$^\textrm{\scriptsize 93}$,    
S.H.~Abidi$^\textrm{\scriptsize 167}$,    
O.S.~AbouZeid$^\textrm{\scriptsize 40}$,    
N.L.~Abraham$^\textrm{\scriptsize 156}$
% ...
\bigskip
\\

% ...
$^{13}$Institute of Physics, Azerbaijan Academy of Sciences, Baku; Azerbaijan.\\
% ...
$^{40}$Niels Bohr Institute, University of Copenhagen, Copenhagen; Denmark.\\
% ...
$^{53}$II. Physikalisches Institut, Georg-August-Universit\"{a}t G\"ottingen, G\"ottingen; Germany.\\
% ...
$^{70}$$^{(a)}$INFN Sezione di Pavia;$^{(b)}$Dipartimento di Fisica, Universit\`a di Pavia, Pavia; Italy.\\
% ...
$^{93}$Department of Physics, Royal Holloway University of London, Egham; United Kingdom.\\
% ...
$^{101}$CPPM, Aix-Marseille Universit\'e, CNRS/IN2P3, Marseille; France.\\
$^{102}$Department of Physics, University of Massachusetts, Amherst MA; United States of America.\\
% ...
$^{128}$Homer L. Dodge Department of Physics and Astronomy, University of Oklahoma, Norman OK; United States of America.\\
% ...
$^{156}$Department of Physics and Astronomy, University of Sussex, Brighton; United Kingdom.\\
% ...
$^{167}$Department of Physics, University of Toronto, Toronto ON; Canada.\\
% ...
$^{a}$ Also at Centre for High Performance Computing, CSIR Campus, Rosebank, Cape Town; South Africa.\\
$^{b}$ Also at CERN, Geneva; Switzerland.\\
% ...
$^{*}$ Deceased

\end{flushleft}


\end{document}